\documentclass[
twocolumn,
superscriptaddress,
showpacs,
nofootinbib,
nobibnotes,
amsmath,amssymb,
aps,
pra,
floatfix,
]{revtex4}

\usepackage{graphicx}
\usepackage{dcolumn}
\usepackage{bm}
\usepackage{natbib}
\usepackage[caption=false]{subfig}
\usepackage{color}
\newcommand{\ket}[1]{\left\vert#1\right\rangle}

\newcommand{\abs}[1]{\left\vert#1\right\vert}

\def \be {\begin{equation}}
\def \ee {\end{equation}}

\begin{document}

\preprint{APS/123-QED}

%
\title{Discrete time quantum walk with nitrogen-vacancy centers\\ in diamond coupled to a superconducting flux qubit}
%
\author{Ali \"U. C. Hardal}
\affiliation{Department of Physics, Ko\c{c} University, \.Istanbul, 34450, Turkey}
\affiliation{Research Center of Integrative Molecular Systems (CIMoS), Institute for Molecular Science, Okazaki, Aichi 444-8585, Japan}
\author{Peng Xue}
\affiliation{Department of Physics, Southeast University, Nanjing 211189, China}
\author{Yutaka Shikano}
\email{yshikano@ims.ac.jp}
\affiliation{Research Center of Integrative Molecular Systems (CIMoS), Institute for Molecular Science, Okazaki, Aichi 444-8585, Japan}
\affiliation{Schmid College of Science and Technology, Chapman University, Orange, CA 92866, USA}
\author{\"{O}zg\"{u}r E. M\"{u}stecapl{\i}o\u{g}lu}
\email{omustecap@ku.edu.tr}
\affiliation{Department of Physics, Ko\c{c} University, \.Istanbul, 34450, Turkey}
\author{Barry C. Sanders}
\affiliation{Institute for Quantum Information Science, University of Calgary, Alberta T2N 1N4, Canada}
\date{\today}
\begin{abstract}
We propose a quantum-electrodynamics scheme for implementing the discrete-time, coined quantum walk with the walker corresponding to the phase degree of freedom for a quasi-magnon field realized in an ensemble of nitrogen-vacancy centres in diamond. The coin is realized as a superconducting flux qubit.  Our scheme improves on an existing proposal for implementing quantum walks in cavity quantum electrodynamics by removing the cumbersome requirement of varying drive-pulse durations according to mean quasiparticle number. Our improvement is relevant to all indirect-coin-flip cavity quantum-electrodynamics realizations of quantum walks. Our numerical analysis shows that this scheme can realize a discrete quantum walk under realistic conditions.
\end{abstract}

\maketitle
\section{\label{sec:intro}Introduction}
Spin ensembles that are coupled to quantum electrodynamics (QED) systems and solid-state
qubits are promising hybrid architectures~\cite{lloyd} for quantum memories~\cite{qmemory} and for distributed quantum
networking~\cite{ozgurPRA}. Nitrogen-vacancy (NV) centers
in diamond are outstanding  among the solid-state spin ensembles~\cite{qcSolidStateSpins}
due to their long spin relaxation and dephasing times, scalability~\cite{toyli2010} and existence of
well-established techniques for their manipulation~\cite{bernien,gruber1997,yutaka}.
NV center quantum memory has recently been demonstrated~\cite{NVqmemoryExp}.

Quantum memories require fast and reliable quantum state transfer, which is
closely related to the quantum walk (QW) problem \cite{kempe2003,qwReview,ambainis2003}. In particular,
the discrete time QW (DTQW)~\cite{shikano,kitagawa_rev,qwReview,shikano_rev}
has been highlighted for universal quantum computation \cite{lovett2010} and for quantum state
transfer~\cite{kurzynski2011}. NV centers that are
coupled to superconducting transmission line resonators are considered before for continuous-time QW~\cite{ozgurCTQW}.
Here we propose a scheme for implementing DTQW with NV centers.

Our scheme is based upon the proposals for DTQW in phase space introduced for
cavity QED (CQED) \cite{CQEDqw,agarwal2005,xue2008} and
circuit QED (cirQED) \cite{cirQEDqw}. Like some other intriguing proposals~\cite{lehman2011,becQW,oamLight},
QWs in CQED and cirQED schemes have not been demonstrated yet; despite some successful realizations in different systems \cite{mataloni2012,matjeschk2009,du2003,perets2008,broome2010,kitagawa2012}.
Our procedure could lead to the first solid-state realization of the proposed DTQW in phase space.

We consider an NV center ensemble coupled to a superconducting flux qubit~\cite{marcos2010}, which plays the role of the coin in DTQW.
Weak excitations of NV centers can be described in terms of bosonic quasiparticles. We take into account all the
crystallographic classes  of the NV centers so that the single mode
coupled to the qubit becomes an effective privileged mode, which is a particular superposition of all the modes corresponding to
the quasiparticle excitations in four different NV center crystallographic classes. We derive an exact bilinear spin-boson model that
reduces to Jaynes-Cummings model (JCM), same with that in CQED or cirQED settings, under parametric approximation.

We employ the similar but improved DTQW scheme proposed for JCM to our case, by removing the cumbersome requirement
of varying drive-pulse durations. This allows for the direct realization of DTQW with regular time steps and simplifies the
implementation in all indirect-coin-flip cavity QED realizations of QW including our full solid-state proposal.

This work is organized as follows. Section~\ref{sec:model} presents the model Hamiltonian, describing the
coupling of the NV center excitations to the flux qubit.
The QW Hamiltonian is obtained by employing canonical transformations. Section~\ref{sec:results} describes how the model Hamiltonian is used
to evolve the initial state of the QW using the quantum master equation by taking into account the decay channels from the
dephasing and relaxation of the qubit as well as the relaxation of the NV centers. The dynamics of the mean number of NV center quasiparticles (called  quasimagnons), the populations of the qubit levels representing the coin operation,
and the Holevo standard deviation are reported. The phase distribution and the
Wigner function plots are also provided to reveal details of the QW
dynamics in the quasimagnon phase space.
A linear fit analysis for the Holevo measure is presented and used to distinguish a QW from the classical
random walk evolution. Section~\ref{sec:con} is devoted to the summary and conclusions.
\section{Model System}\label{sec:model}
\subsection{NV center ensemble Hamiltonian}
A theoretical model of NV centers in diamond can be developed based on the six-electron description of the $^3A$
electron spin triplet ground state of NV$^-$ \cite{nvGround1,nvGround2}. The
hyperfine ($\sim 2$ MHz for $^{14}$N), quadrupole ($\sim 5$ MHz for $^{14}$N), and strain ($<10$ MHz)  interactions are negligible \cite{reviewNV} relative to zero-field splitting $\sim 2.88$ GHz. Similar conclusion is applicable to $^{13}$C as well~\cite{smeltzer2011}.
The Hamiltonian of a single NV center in the absence of external electric and magnetic fields can be written as~\cite{loubser1978,doherty2012}
\begin{eqnarray}
H=\tilde S^T\tilde D\tilde S,
\end{eqnarray}
where $\tilde S$ is the spin-1 matrix and $\tilde D$ is the zero-field splitting matrix.
The NV center Hamiltonian in the principal axes of the molecular body frame becomes diagonal so that
\begin{eqnarray}
H=D\left[S_z^2-\frac{1}{3}S(S+1)\right],
\end{eqnarray}
where $S=1$ and $D/2\pi\approx 2.88$ GHz \cite{manson2006}.
By taking the quantization axis $z$ as the alignment of the NV center, the Hamiltonian can be expressed in the spin basis
$|S,m\rangle\equiv|m\rangle$ as
\begin{eqnarray}
H=D\sum_{m=\pm 1,0}m^2|m\rangle\langle m|,
\end{eqnarray}
where we dropped the constant term.

There are four possible quantization axes in the NV center, which are all equivalent. We need to take them
into account when we consider an ensemble of NV centers. Let us denote these axes by $f=1, \dots, 4$ and introduce collective
operators, which are analogs of Hubbard operators for strongly correlated electron systems \cite{HubbardOp}, such that
\begin{eqnarray}\label{eq:hubbard}
X_f^{mn}=\sum_{j=1}^{N_f}|m\rangle_j^{(f)}\langle n|_j^{(f)}.
\end{eqnarray}
Here, $\mid m\rangle^{f}_j$ is the spin state of the $j$-th NV center aligned along the crystallographic (molecular) axis-$f$ which
is taken as the quantization axis.
$N_f$ denote the number of NV centers aligned along the quantization axis $f$. The Hamiltonian for the NV center ensemble becomes
\begin{eqnarray}
H_{\text{NV}}=D\sum_{mf}m^2X_f^{mm}.
\end{eqnarray}
\subsection{Flux qubit Hamiltonian}
A flux qubit consists of three or more superconducting Josephson junctions, one of which is smaller than the others
\cite{orlando1999, mooij1999}.
The phase quantization condition for the
superconducting wave functions around
the qubit loop reduces the system to a two-dimensional phase space, where the qubit can be effectively considered
as a two-particle system subject to a double-well potential  \cite{polThesis}. A two-level approximation can be made to the band
structure of the potential by assuming that the external flux is in the vicinity of half of the  flux quantum, where the two lowest bands are closest
to each other. A small overlap of the wave functions of the two levels opens the qubit gap $\Delta_q$, which
can be tuned \cite{paauw2009,fedorov2010,zhu2010} by replacing the smallest Josephson junction with two parallel junctions, a DC-SQUID.
On the basis of the two counter-rotating persistent currents in the qubit loop, the qubit Hamiltonian can be written as
\begin{eqnarray}
H_q = \frac{1}{2}\left( \Delta_q \sigma_x+\epsilon_q \sigma_z\right),
\end{eqnarray}
where $\epsilon_q$  is the magnetic energy bias of the qubit that contributes away from the symmetry point, when there is a net persistent current in the qubit loop.

Pauli spin operators $\sigma_{x,y,z}$ can be conveniently transformed from the bare current basis to the dressed current basis by $U=\exp{(-i\theta\sigma_y)}$
with $\theta=\arctan{(\Delta_q/\epsilon_q)}$. The Hamiltonian becomes
\begin{eqnarray}
H_q=\frac{\omega_q}{2}\sigma_z,
\end{eqnarray}
with the qubit frequency $\omega_q=\sqrt{(\Delta_q^2+\epsilon_q^2)}/2$.
\subsection{Flux qubit-NV ensemble coupling}
We assume the same conditions as in a recent experiment~\cite{zhu2011}, wherein the flux qubit was coupled to the NV center ensemble located at the top of the central region of the flux qubit loop.
The dimensions of the ensemble are sufficiently small for the spatial variations in the magnetic field over the ensemble to be ignored.
In such a case, the magnetic fields generated by the supercurrents are expected to interact with the ensemble homogeneously. In
the persistent current basis the electronic Zeeman interaction is given by
\begin{eqnarray}
H_{\text{int}}=\sum_{\alpha=x,y,z}g^f_{\alpha}S^{\alpha}_f\sigma_z,
\end{eqnarray}
where $S^{\alpha}_f$ are the total spin-1 operators for the NV ensemble that are related to the generators
of the SU(3). Nuclear Zeeman interaction is three orders of magnitude smaller and neglected. Here, $\alpha$  denotes
the coordinate in the molecular body frame; and $\sigma_z$ is associated with the magnetic field in the persistent current basis.

\begin{figure}[!ht]
\begin{center}
\includegraphics[width=0.4\textwidth]{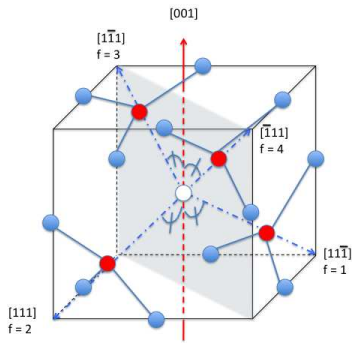}
\caption{\label{fig1} (Color online) Four equivalent crystallographic groups of the NV centers. Symmetry axes $\langle 111 \rangle$ make the same angle with the magnetic field parallel to the axis directed along $[001]$. The blue spheres along the sides of the cube represent the C atoms; the white sphere at the center represents the vacancy; the red spheres along the $\langle 111 \rangle$ axes represent the N atoms. We use the short hand notation $f=1,2,3,4$ for the axes $[11\bar 1],[111],[1\bar 1 1],[\bar 111]$, respectively. }
\end{center}
\end{figure}

The electronic Zeeman interaction is diagonal in the molecular body frame. The symmetry axes 
$\langle 111\rangle=[11\bar 1],[111],[1\bar 1 1], [\bar 1 11]$, shown in Fig.~\ref{fig1},
which correspond to four crystalographic classes $f=1,2,3,4$, respectively, are taken
to be the $z$ axes of the molecular body frames. The coupling strengths $g^f_\alpha=2g_e\mu_BB_{f\alpha}$, where $g_e$ is the electronic Land\'e g factor, $\mu_B$ is the Bohr magneton and
$B_{f\alpha}$ is the $\alpha$-component of the magnetic field acting in the molecular frame of the
crystallographic class $f$, can be adjusted by the orientation of the diamond sample. In a typical experimental
setting~\cite{zhu2011}, the crystal axis $[001]$ is parallel to the magnetic field.
The crystal surface $(001)$ is perpendicular to the magnetic field and placed on top of the flux qubit. 

If we take the $[001]$ direction
as the crystal lab frame $z$ axis, then the Zeeman interaction coefficients become proportional to the direction cosines of the
lab and body frame $z$ axes. The angle between the $f=1$ and $f=2$ axes and the angle between the $f=3$ and $f=4$
axes are the same in the diamond unit cell. 
The magnetic field axis bisects them and thus the magnetic field makes the same angle of $54.7^\circ$
with the every body frame $z$ axes. The molecular body frames are chosen such that the $y$ axes are perpendicular to the magnetic field and the $x$ axes make an angle of $35.3^\circ$ with the magnetic field. This makes
$g^f_{\alpha}=g_{\alpha}$ the same for all
crystallographic classes $f$ and  $g_{z}$, associated with the smaller direction cosine,
becomes smaller relative to $g_x$ while $g_y=0$~\cite{marcos2010,zhu2011}. These lead to the reduced interaction of the form
\begin{eqnarray}
H_{\text{int}}=&\sum_f  \sigma_z [g_x(X_f^{-0}+X_f^{+0}+\text{H.c.})\nonumber\\
&+g_z(X_f^{++}-X_f^{--})],
\end{eqnarray}
where H.c. stands for Hermitian conjugate. Here we replaced the collective spin operator by the
analogs of the Hubbard operators introduced in Eq.~(\ref{eq:hubbard}) such that $S_x^f=X_f^{-0}+X_f^{+0}+X_f^{0-}+X_f^{0+}$ and $S_z^f=X_f^{++}-X_f^{--}$.

We shall proceed under the
assumptions of weak excitations and a large number of spins in the ensemble. Experiments can achieve coupling of
$N\sim 3.1\times 10^7$ NV centers directly above the flux qubit \cite{zhu2011}. This number is sufficiently large for
our weak-excitation conditions. We consider about $5-10$ excitations in our simulations.
Under these conditions, operators $X_f^{mn}$ can be conveniently related to bosonic operators $x_{fm}$ such that
\begin{eqnarray}
X_f^{mn}=x_{fm}^\dag x_{fn},
\end{eqnarray}
in terms of which the corresponding terms in the Hamiltonian $H=H_{\text{NV}}+H_{q}+H_{\text{int}}$ becomes
\begin{eqnarray}
H_{\text{NV}}&=&D\sum_{mf}m^2x_{fm}^\dag x_{fm},\\
H_{\text{int}}&=&\sum_f g \sigma_z (x_{f-}^\dag x_{f0}+x_{f+}^\dag x_{f0}+ \text{H.c.}),
\label{eq:Hint_x_op}
\end{eqnarray}
where we neglected $g_z(x_{f+}^\dag x_{f+}-x_{f-}^\dag x_{f-})$ and redefined $g_x:= g$ in Eq.~(\ref{eq:Hint_x_op}). The operators $x_{fm}$ play the role of Schwinger bosons of $SU(N_f)$
subject to the constaint $\sum_m \langle x_{fm}^\dag x_{fm}\rangle=N_f$ \cite{li2004,shen2002,bauer2012}.
Schwinger bilinear boson representation of the Hubbard operators preserve their algebra.
The algebraic relations are summarized in Appendix~\ref{appendix:HubbardAlgebra}.

The quasiparticles of NV center excitations are analogous
to quantum particles associated with the spin waves, and hence, we call them quasimagnons.
We introduce quasimagnon operators as follows:
$x_{f+}\equiv a_f,\quad x_{f-}\equiv b_f$,
and proceed by replacing $x_{f0}\rightarrow \sqrt{N_f}$ which is analogous to the parametric pump approximation
in quantum optics \cite{parametricPump1}.
Parametric approximation has quite a wide range of validity, which is especially enhanced for initially
coherent states \cite{parametricPump2}. To the first order the error is as large as the
of the square root of the mean quasimagnon number. Fractional error, relative to the mean number
of quasimagnons, is about the size of relative fluctuation which is negligible in our case $\sqrt{N}/N\sim 0.01$.
The change from the algebra preserving Schwinger bosons to quasimagnons by the parametric aproximation is equivalent to
deforming the Holstein-Primakoff map under group contraction by large total spin approximation~\cite{holstein1940}.

We consider initially coherent spin ensembles for our quantum walk procedure as well, which could make
our procedure potentially applicable for higher number of excitations, or equivalently for smaller
number of NV centers coupled to the flux qubit. A systematic study of how tolerant our procedure to smaller number of
NV centers however needs further examinations without parametric approximation, which is beyond the scope of our
present investigations.

The Hamiltonian of the NV centers then reduces to
\begin{eqnarray}
H_{\text{NV}}=\sum_f D(a_f^\dag a_f+ b_f^\dag b_f).
\end{eqnarray}
We interpret the presence of two quasimagnon operators by associating them with quasispin waves with two possible polarizations. The final Hamiltonian is then an eight-mode model, in which there are four quasimagnon modes, each with two different polarization degrees of freedom.

By employing the rotating wave approximation for $g\ll D,\omega_q$, transforming to the dressed current basis at the qubit symmetry point,
and using the quasimagnon picture, we can express the interaction as
\begin{eqnarray}
H_{\text{int}}=\sum_f [\eta_f(\sigma_+(a_f+b_f)+\sigma_-(a_f^\dag+b_f^\dag)],
\end{eqnarray}
with $\eta_f=g\sqrt{N_f}$ being the collective coupling strengths for the quasimagnons and flux qubit system.
Here $\sigma_{\pm}=\sigma_x\pm\mathrm{i}\sigma_y$ are the Pauli spin ladder operators.

This Hamiltonian can be simplified by distinguishing sets of  ``coupled modes" and ``uncoupled modes," analogous to Morris-Shore transformations \cite{morris1983},
which are combinations of quasimagnon modes
such that they are respectively interacting and noninteracting with the flux qubit. We list these modes in Appendix~\ref{appendix:quasimagnonModes}. Eventually, only a single effective privileged
mode becomes coupled to the qubit. The annihilation operator of that mode is given by
\begin{eqnarray}\label{eq:quasimagnonMode}
c=\frac{1}{\sqrt{2N}}\sum_f \sqrt{N_f}(a_f+b_f).
\end{eqnarray}
The interaction Hamiltonian is then simplified to the usual Jaynes-Cummings model, though the single privileged (bright)
mode is actually a spesific superposition of all the quasimagnon modes,
\begin{eqnarray}
H_{\text{int}}=\eta(\sigma_+c+\sigma_- c^\dag),
\end{eqnarray}
where the collective spin-qubit coupling strength is $\eta=\sqrt{2N}g$.
We ignore the effects of inhomogenous broadening and spatial inhomogeneity here and provide a brief discussion in  Appendix~\ref{appendix:inhomogenity}.
\subsection{Quantum walk Hamiltonian}
We introduce an additional term that represents the driving field on the diamond sample
\begin{eqnarray}
H_d=\epsilon(t)(c e^{i\omega_d t} + c^\dag e^{-i\omega_d t}),
\end{eqnarray}
where $\omega_d$ is the drive frequency and $\epsilon(t)$ is the time-dependent amplitude of the drive pulse.
Consistent with our spatial homogenity of the interaction assumption, the drive field is coupled with the quasimagnon mode
defined in Eq.~(\ref{eq:quasimagnonMode}).
We assume that the drive is of square-wave form with pulse duration $t_H$ and amplitude $\epsilon_0$.
The total Hamiltonian of the qubit ensemble becomes the Jaynes-Cummings model Hamiltonian such that
\begin{eqnarray}
H&=&\frac{\omega_q}{2}\sigma_z+Dc^\dag c+\eta(\sigma_+c+\sigma_- c^\dag)+H_d(t)
.\end{eqnarray}

This Hamiltonian is  the same as that proposed for a QW in cirQED, although in our case,
the role played by the microwave photons is assumed by the quasimagnons. Here,
the qubit-NV center ensemble coupling $\eta$ is about $\eta/2\pi\sim 70$ MHz \cite{zhu2011}. In contrast to spontaneous decay and the dephasing of the qubit,
the qubit-NV center ensemble interaction is in the strong-coupling regime. By employing a rotating frame transformation
$H_R=U^\dag H^\prime U-\Lambda$ with $U=\exp{(-i\Lambda t)}$, where we take $\Lambda=\omega_d(c^\dag c
+\sigma_z/2)$, the Hamiltonian can be written in the form
\begin{eqnarray}\label{eq:Hamiltonian_timeIndep}
H=-\delta_c c^\dag c-\delta_d\frac{\sigma_z}{2}+\eta(c^\dag \sigma_-+c\sigma_+)+\epsilon(c+c^\dag).
\end{eqnarray}
Here, $\delta_d=\omega_d-\omega_q$ and $\delta_c=\omega_d-D$ are introduced.

The detuning between the qubit and the NV center
ensemble is $\Delta=\delta_d-\delta_c=D-\omega_q$, which can be adjusted by $\omega_q$ to satisfy $\Delta\gg\eta$ so that the
Fr\"ochlich canonical transformation \cite{frohlich1950,schrieffer1966,nakajima1955} $H^\prime=\exp{(-S)}H\exp{(S)}$, with the transformation operator
\begin{eqnarray}
S=\frac{\eta}{\Delta}(\sigma_- c^\dag-\sigma_+ c),
\end{eqnarray}
can be applied to obtain an effective Hamiltonian
\begin{eqnarray}\label{eq:effectiveH}
H_{\text{eff}}&=&\chi c^\dag c\sigma_z+\frac{\Omega_R}{2}\sigma_x
+\epsilon(c+c^\dag)-\delta_cc^\dag c-\frac{\delta_d}{2}\sigma_z \nonumber\\
&+&\frac{\chi}{2}\left(1_2+\sigma_z\right),
\end{eqnarray}
where $1_2$ is the $2\times 2$ unit matrix; and the parameters
\begin{eqnarray}
\chi=\frac{\eta^2}{\Delta},\quad\Omega_R=\frac{2\eta\epsilon(t)}{\Delta}
\end{eqnarray}
are associated with the conditional rotational shift and the Hadamard coin terms in the Hamiltonian. The Hadamard coin term
only operates when the pulse is on, whereas the shift operation is constantly active. The conditional shift operator
$R(\chi):=\exp{(\mathrm{i}\chi c^\dag c\sigma_z)}$ rotates the phase states 
\begin{eqnarray}\label{eq:phaseState}
\ket\phi=\frac{1}{\sqrt{M}}\sum_{n=0}^{M-1}
{\mathrm e}^{\mathrm{i} n\phi}\ket n,
\end{eqnarray}
on the phase space such that $R(\chi)\ket\phi\ket\sigma_z=\ket{\phi+\sigma\chi}\ket\sigma_z$. The rotation can be
clockwise or counterclockwise depending on the state of the qubit $\ket{\sigma=\pm 1}_z$. Here $\ket n$ are the Fock states in an Hilbert space of
dimension $M$. The conditional rotation is restricted to a cycle whose radius is determined by the mean quasimagnon number 
$\langle n_c\rangle=\langle c^\dag c\rangle$ and the phase
space coordinates correspond to the mean quadratures $\langle c+c^\dag\rangle/2$ and $\langle(c-c^\dag)\rangle/2\mathrm{i}$. 
The effect of the conditional rotation operator on a coherent state $\ket\alpha$ 
is similar; it translates the phase $\theta_\alpha$ of the coherence parameter $\alpha=\abs\alpha\exp{(\mathrm{i}\theta_\alpha)}$
to $\theta_\alpha\pm\chi$.

We note that in the QW proposal for cirQED, the parameter $\Omega_R$ is found to depend on $\omega_d$ by the relation
$\Omega_R=2\eta\epsilon(t)/\delta_c$. The dependence of  $\Omega_R$ on the drive frequency results in a QW with
varying step sizes in time, and complicated pre-adjustments in pulse duration are therefore required. Here, we do not have a
drive-frequency-dependent coin operation and we have a fixed step size as well as fixed pulse duration. This makes the implementation
of the QW more straightforward.  In contrast to our approach described here, a displaced frame transformation relative to the resonator field was previously used in the cirQED model of a
bit-flip gate \cite{blais2007}, in addition to the Fr\"ochlich
transformation. This lead to vanishingly small cavity photon numbers and negligible ac Stark shifts. In our approach, we do not use a displaced frame, which allows for the presence of a larger number of quasimagnons.
\section{Results and Discussion}\label{sec:results}
The dynamics of the effective Hamiltonian is subject to decoherence processes. By denoting
the qubit dephasing rate as $\gamma_\phi$, the qubit relaxation rate as $\gamma_1$, and the decay
of the NV center excitations as $\Gamma$, we can write the quantum master equation of the system,
under the Markovian and Born-Markov approximations, which are suitable for our weakly coupled system \cite{gardiner}, as
\begin{eqnarray}{\label{eq:master}}
\dot\rho=-i[H,\rho]+\Gamma{\cal D}[c]\rho+\gamma_1{\cal D}[\sigma_-]\rho
+\frac{\gamma_\phi}{2}{\cal D}[\sigma_z]\rho.
\end{eqnarray}
Decoherence processes associated with an operator $x=c,\sigma_-,\sigma_z$ are described by the Liouvillian superoperators
${\cal D}[x]\rho=(2x\rho x^\dag-x^\dag x\rho-\rho x^\dag x)/2$ in Lindblad form.
We introduced the effective Hamiltonian to illustrate the underlying QW dynamics,
but obtain quantitative
results from the Hamiltonian of the system given in Eq.~(\ref{eq:Hamiltonian_timeIndep}).

The effective Hamiltonian is in
We do not simulate
the effective Hamiltonian to avoid further approximations. The effective Hamiltonian only serves to interpret and design the dynamics of the system.

We solve the master equation using the QuTip package \cite{qutip} in \textsc{python} software. The initial condition is
considered to be
\begin{eqnarray}
|\psi(0)\rangle=|\alpha\rangle(|g\rangle+i|e\rangle)/\sqrt{2},
\end{eqnarray}
where the quasimagnons are in a coherent state with amplitude $\alpha$ and the qubit is in a superposition of its ground and excited states. It is assumed that excitations of the
NV center ensemble is initially prepared in
a coherent state. We consider a Fock space of quasimagnons with the dimension $17,$ which is sufficiently large to accommodate the selected $\alpha=3$ in the numerical analysis.
A classical magnetic field pulse can be used to prepare the quasimagnon coherent state analogous to the typical driven oscillator case \cite{shumovsky1995}, as for large number of NV centers the atomic coherent states converges to the bosonic coherent states \cite{mandelWolf}.
The modes to be included in the initial coherent state could be selected by external bias magnetic fields.
The initial coherent superposition of the flux qubit levels can be generated by a resonant magnetic pulse of area $\pi$.

Most of the parameters we use in our simulations can be taken as the same values with those used in the
original circuit QED and cavity QED proposals \cite{cirQEDqw,xue2008} except $\gamma_1,\Gamma,D$ and $\epsilon_0$.
Flux qubit has a wide range of tunability and we set its gap frequency as $\omega_q/2\pi=7$ GHz \cite{mooij1999}.
Recently developed flux qubits can have dephasing times in the order of few
microseconds \cite{steffen2010} and
we set  $\gamma_\phi/2\pi=0.31$ MHz.
Recent experiments allow for direct coupling between the flux qubits and the NV centers with strengths of about $\sim 70$ MHz \cite{zhu2011}. Number of NV centers and the flux qubit size can be used to control the strength of the
coupling, which is taken here as $\eta/2\pi=100$ MHz.

\begin{figure*}[!t]
\subfloat[][]{
\includegraphics[width=0.4\textwidth]{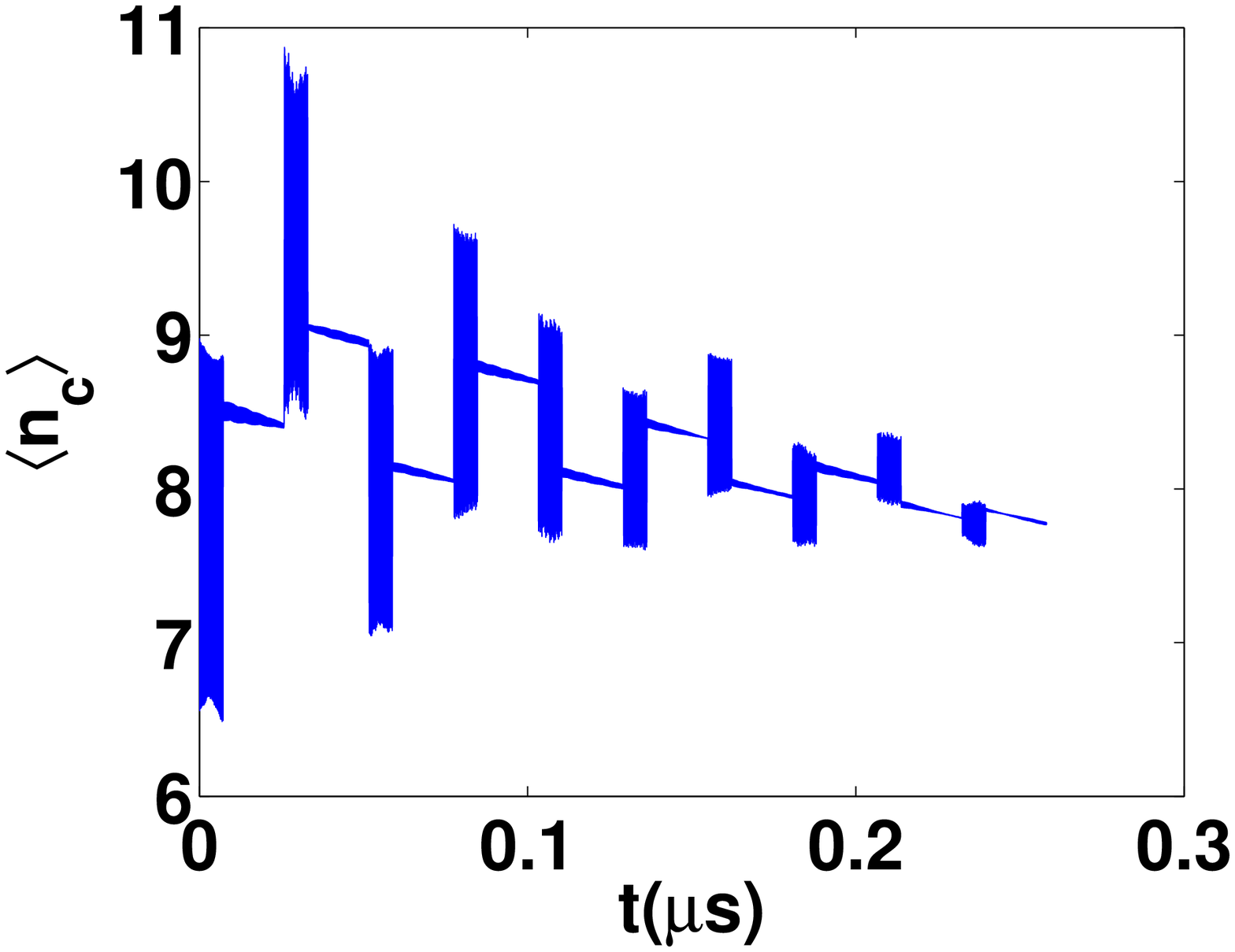}
\label{fig2a}}
\subfloat[][]{
\includegraphics[width=0.4\textwidth]{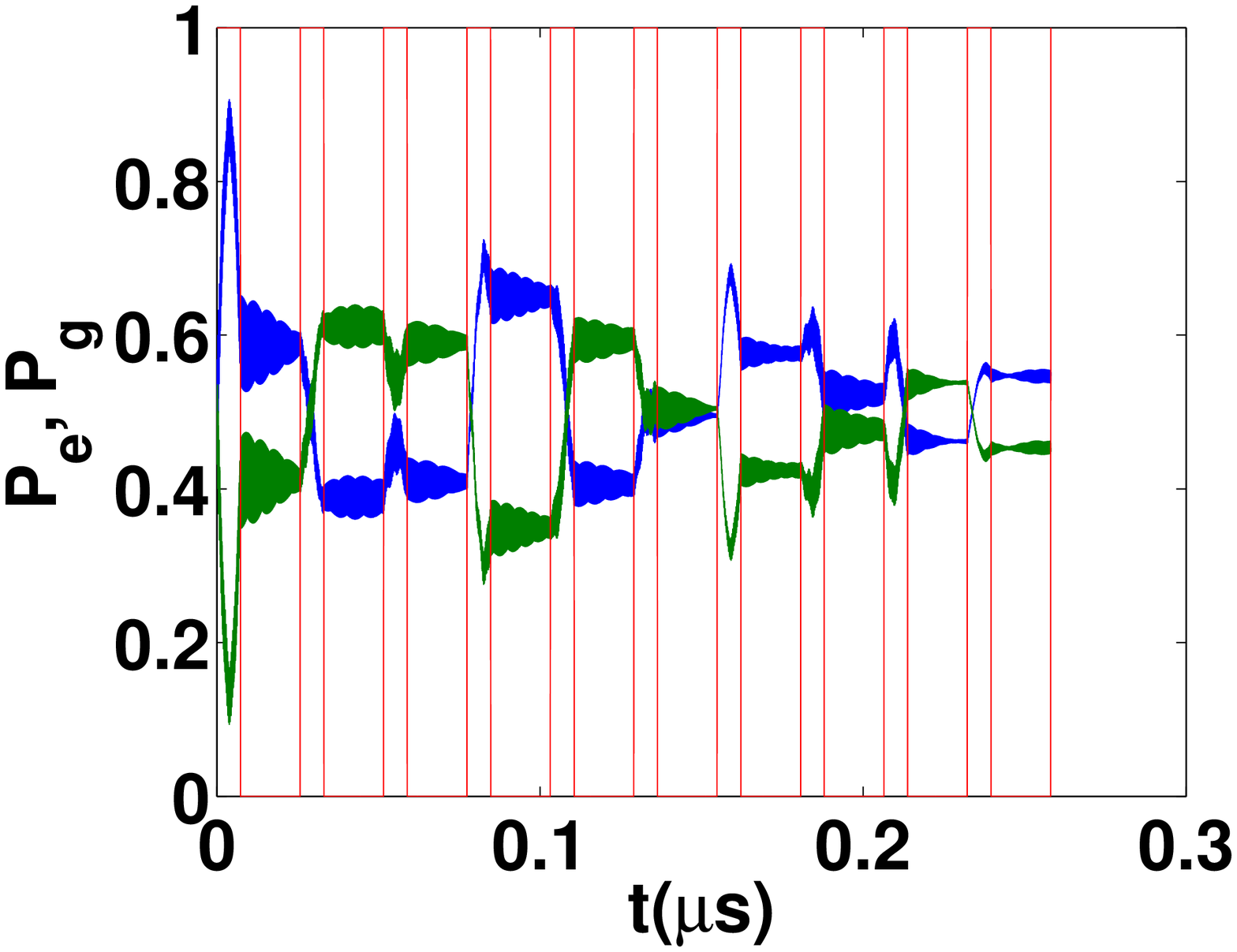}
\label{fig2b}}
\caption{\label{fig2} (Color online) (a)
Time (t) dependence of the mean number of quasimagnons $\langle n_c \rangle$ and (b) of the qubit-level populations $P_{e,g}$
for the initial state
$|\psi(0)\rangle=|\alpha=3\rangle(|g\rangle+i|e\rangle)/\sqrt{2}$.
$\langle n_c \rangle$ fluctuates in time, mainly when the pulse is on (coin toss).
Regular intervals of the pulse in (b) indicate the regular discrete time steps of the QW.
The parameters used in our simulations are
$\omega_q/2\pi=7$ GHz, $D/2\pi=2.87$ GHz,
$\eta/2\pi=100$ MHz, $\epsilon_0/2\pi=1$ GHz, $\gamma_1/2\pi=0.02$ MHz,
$\gamma_\phi/2\pi=0.31$ MHz, and $\Gamma/2\pi=0.1$ MHz. The repetition rate of the driving pulsed
laser is taken as $\omega_p=\chi d$, with $d=16$ being the
number of sites on the cycle of a QW.}
\end{figure*}

As our improved procedure is general and applicable to previous circuit QED or cavity QED settings \cite{cirQEDqw,xue2008}, we shall first use the original values $\gamma_1/2\pi=0.02$ MHz and $\epsilon_0/2\pi=1$ GHz for comparison of our improved procedure with the original ones in Refs. \cite{cirQEDqw,xue2008}. The minor difference of no consequence is that we use $D/2\pi=2.87$ GHz \cite{manson2006} for the
 zero-field splitting between the $\mid m=0\rangle$ and
$\mid m=\pm 1\rangle$ ground state spin triplet levels of the NV center. This corresponds to cavity or transmission line mode
frequency, which is taken as $5$ MHz in Refs. \cite{cirQEDqw,xue2008}.

Following the comparison, we present the results for the actual flux qubit and NV ensemble parameters.
Relaxation rate of the flux qubit is of the same order with its dephasing rate
so that we set $\gamma_1/2\pi=0.31$ MHz \cite{steffen2010}.
Longitudinal relaxation time $T_1$ of the NV center electron spin is in the
order of few milliseconds at room temperature \cite{nvGround2}
and gets longer by reducing the temperature.
The dephasing time of the spin
$T_2$ is in the order of $0.3-2$ ms and gets longer by
increasing the purity of the diamond sample isotropically \cite{balasubramanian2009}. In the case of quasimagnons, the
decoherence rate $\Gamma$ would be collectively enhanced, similar to the collectively enhanced coupling strength. For
$N=3.1\times 10^7$ and $T_2 = 2$ ms, we find $\Gamma/2\pi=2.78$ which is larger than the circuit QED and
cavity QED cases where $\Gamma/2\pi=0.1$ MHz \cite{cirQEDqw,xue2008}. To compensate  it we take larger
drive strength $\epsilon_0/2\pi=10$ GHz.

Relative to the decoherence \cite{yoshihara2006} and dephasing \cite{kakuyanagi2007} of the qubit, as well as to the relaxation of the NV centers, the system is in the strong-coupling regime, though still not in the ultrastrong-coupling regime, which would challenge the validity of rotating wave and Born-Markov approximations.

The repetition rate of the driving pulsed
laser is taken to be $\omega_p=\chi d$, with $d=16$ being the
number of sites on the cycle of a QW. This yields the duration of each time step of the QW as
$t_p=2\pi/\omega_p$. This time covers the pulse duration of $t_H=\pi/(\sqrt{2}\Omega_{R0})$ with $\Omega_{R0}:=2\pi\epsilon_0/\Delta$
as well as the time between pulses. Hadamard coin operation induced by the pulse can be described by the Hamiltonian
\begin{eqnarray}
H_{\text{coin}}=\frac{\Omega_R}{2}(\sigma_x+\sigma_z),
\end{eqnarray}
if we take the operation time as $t_H$; and if we decouple the quasi-magnon
dynamics from the qubit dynamics. For that aim, let us drop the constant term $(\chi/2)1_2$ and rewrite the effective QW Hamiltonian in Eq.~(\ref{eq:effectiveH}) as
\begin{eqnarray}
H_{\text{eff}}=H_{\text{coin}}+(\chi c^\dag c -\bar{n})\sigma_z-
\delta_c c^\dag c +\epsilon (c+c^\dag),
\end{eqnarray}
under the condition of
\begin{eqnarray}
\chi \bar{n}-\frac{\delta_d}{2}+\frac{\chi}{2}=\frac{\Omega_R}{2}.
\end{eqnarray}
This condition can be satisfied by taking the pulse frequency to be
\begin{eqnarray}
\omega_d=(2\bar{n}+1)\chi +\omega_q-\Omega_R,
\end{eqnarray}
where $\bar n=|\alpha|^2$ is the initial number of quasi-magnons. $\epsilon_0$ is the amplitude of the square pulse. If the number fluctuations are sufficiently small, then the dynamics of the quasi-magnons
are uncoupled from the qubit degrees of freedom, whose evolution would be a faithful representation of that of an ideal Hadamard coin. Quantitatively, if we denote the number fluctuations by $\delta n$, the
condition of validity can be expressed as $\chi\delta n\ll\Omega_R/2$ or $\delta n\ll \epsilon_0/\eta$,
which becomes $\delta n\ll 10-100$ for our parameters.
Our simulations indicate that number fluctuations remain in the order of few quasi-magnons; and thus the coin operation can be simulated.
In contrast to the cirQED and CQED QW proposals, we do not change $\omega_d$ by changing $\bar n$ at every step. Furthermore,
the pulse duration, and hence, the  time steps, are fixed in our approach. These two differences
allow for simpler implementation as well as for direct interpretation in terms of DTQW with regular time steps.

We first examine the behavior of the number of quasimagnons, $\langle n_c\rangle=\langle c^\dag c\rangle$, and the qubit excited and ground level populations, which are respectively defined to be $P_e=(1+\langle \sigma_z\rangle)/2$, and $P_g=(1-\langle \sigma_z\rangle)/2$.
Their time evolutions are shown in Fig.~\ref{fig2}.
The initial number of quasimagnons is $n_c(0)=9$. Fig.~\ref{fig2a} indicates that during the course of dynamics
mean number of quasimagnons fluctuates because of the drive and dissipation. Relatively Strong fluctuations in the mean
quasimagnon number occur mainly when the pulse is on, or when the Hadamard coin is tossed.
When the pulse is off, a conditional phase shift operation characterized by the parameter $\chi$ is set into action.
During this rotational QW step in quasimagnon phase space, fluctuations of the $\langle n_c\rangle$ are much smaller.
The dynamics of the coin, or the populations of the qubit levels, can be followed by the
Fig.~\ref{fig2b}. We have marked the
regions during which the pulse is on. The dynamics and exchanges between the upper and lower level populations reveal that the qubit evolves in close correlation with the simulation of a pure Hadamard coin.
Exchanges only occur when the pulse is on, or at the end of the coin toss.

\begin{figure}[!ht]
\begin{center}
\includegraphics[width=0.4\textwidth]{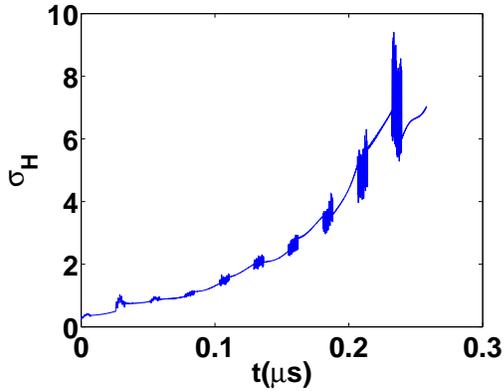}
\caption{\label{fig3} (Color online) Time (t) dependence of the
Holevo standard deviation measure $\sigma_{\text H}$
in the system for the initial state
$|\psi(0)\rangle=|\alpha=3\rangle(|g\rangle+i|e\rangle)/\sqrt{2}$.
Its decrease beyond the seventh step is due to the cyclic
wrapping-up effect in terms of the definition of the Holevo measure;
it does not represent the actual behavior of the QW. The parameters are the same as Fig.~\ref{fig2}.}
\end{center}
\end{figure}
QW in the quasimagnon phase space needs to be characterized by examination of the dispersion of the angular variable $\phi$. For cyclic variables the variance is conveniently determined by the Holevo standard deviation which is defined as \cite{holevo1984}
\begin{eqnarray}
\sigma_{\text{H}}=\sqrt{\abs{\langle\mathrm{e}^{\mathrm{i}\phi}\rangle}^{-2}-1},
\end{eqnarray}
where
\begin{eqnarray}
\abs{\langle\mathrm{e}^{\mathrm{i}\phi}\rangle}=
\left|  \int_{-\pi}^{\pi}{\mathrm d}\phi P(\phi) {\mathrm e}^{\mathrm{i}\phi}\right|
\end{eqnarray}
is the {\it sharpness} of the phase distribution. Sharpness is bounded between $0$, for spiked distribution at the mean phase, and $1$,
for a flat distribution. It can be interpreted as the success or fidelity of the phase distribution to estimate the mean phase \cite{berry2000,bagan2008}.
Here we take the mean phase as $0$ and choose
the domain as $[-\pi,\pi]$. The phase distribution $P(\phi)$ is determined from the reduced
density matrix of the quasimagnons $\rho_\text{m}$ by using
\begin{eqnarray}
P(\phi)=\lim_ {M\rightarrow\infty}
\langle \phi\mid\rho_\text{m}\mid\phi\rangle,
\end{eqnarray}
where the phase states $\ket\phi$ in an $M$-dimensional Fock space are defined in Eq.~(\ref{eq:phaseState}).
The quasimagnon reduced density matrix is calculated by,
$\rho_\text{m}=\text{Tr}_{\text{q}}(\rho)$, taking the trace over the qubit degrees
of freedom of the density matrix of the system $\rho$,
which is found by solving Eq.~(\ref{eq:master}). If the phase variance $(\Delta\phi)^2=
\langle \phi^2\rangle-\langle\phi\rangle^2$ is small ($(\Delta\phi)^2 \ll 1$) then
$\sigma_{\text{H}}$ becomes equivalent to the $\Delta\phi$. In contrast to localized distributions,
Holevo variance is infinitely large for flat distributions with zero sharpness.
In our case, this situation corresponds to
large phase diffusion due to decoherence in the system. We examine the Wigner function
evolution to determine the number of steps for which the Holevo measure can be used
for reliable information on the phase variance.

\begin{figure}[t]
\begin{center}
\includegraphics[width=0.45\textwidth]{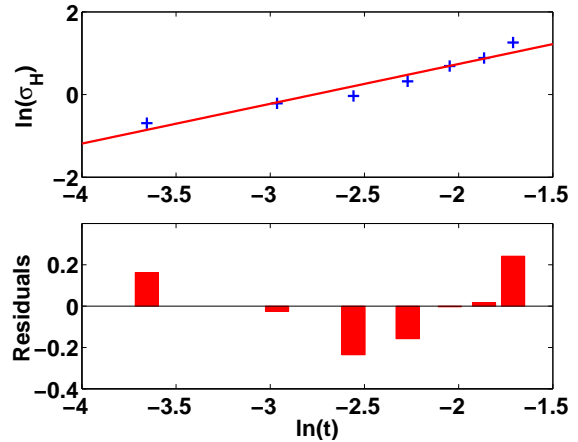}
\caption{\label{fig4} (Color online)
A linear fit to the Holevo measure $\sigma_{\text H}$ in Fig.~\ref{fig3}. A logarithmic scale is
used for both axes. The initial starting point in time is taken
as zero and not used on the logarithmic scale. The slope of the
line is calculated as $\sim \negthickspace0.96$, which is in the ballistic regime and higher than the corresponding
value of $0.5$ for a classical random walk. The parameters are the same as Fig.~\ref{fig2}, except that the time $t$
here is dimensionless, scaled by $1\,\mu$s.}
\end{center}
\end{figure}

Fig.~\ref{fig3} shows that $\sigma_{\text{H}}$ increases in an almost smooth manner during the conditional phase shifts, whereas during the coin toss operation, it shows rapid
oscillations. At the eighth step, more than half of the cycle is covered by the Wigner function (see Fig. \ref{fig6}).
The Holevo measure becomes unsuitable for describing the second order phase variance beyond this point.
Similar behavior is  encountered in the cirQED QW proposal~\cite{xue2008}.
To check whether the phase variance is faster than that of the classical random walk, we employ a linear fit analysis for the first $7$ steps using \textsc{MATLAB}. The result is shown in Fig.~\ref{fig4}.
The upper panel in Fig.~\ref{fig4} reveals that there is a linear relation  between the number of regular time steps and the logarithm of the Holevo standard deviation.
The slope is determined to be $\sim \negthickspace 0.96\pm 0.11$, where $0.11$ is the
linear-regression standard deviation.
This result is close enough to the ballistic case to declare it to be ballistic given that
phase diffusion due to decoherence and wrapping due to periodicity of phase
have occured. In the case of the classical random walk, the slope would be $0.5$.

\begin{figure}[t]
\subfloat[][]{
\includegraphics[width=0.24\textwidth]{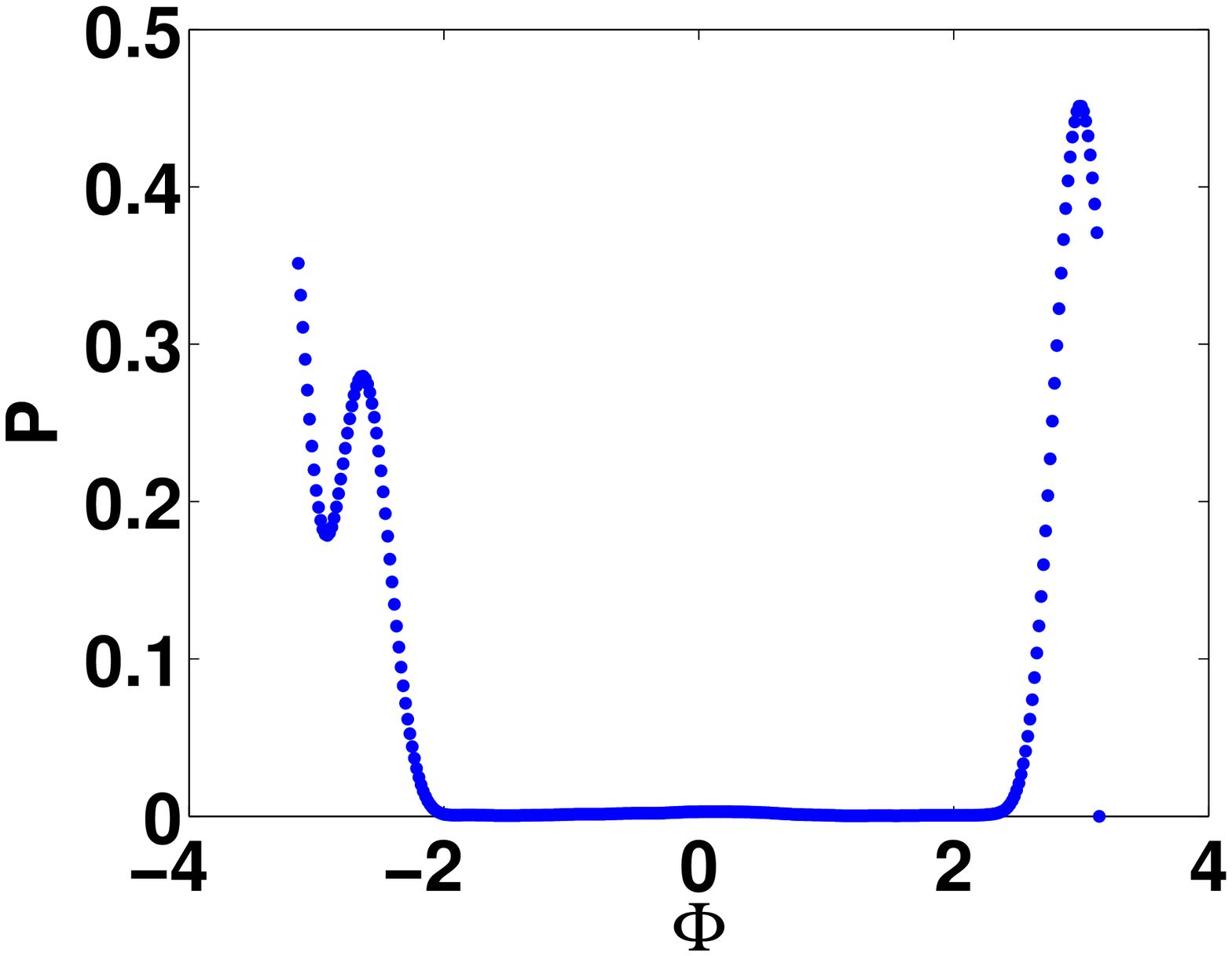}
\label{fig5a}}
\subfloat[][]{
\includegraphics[width=0.24\textwidth]{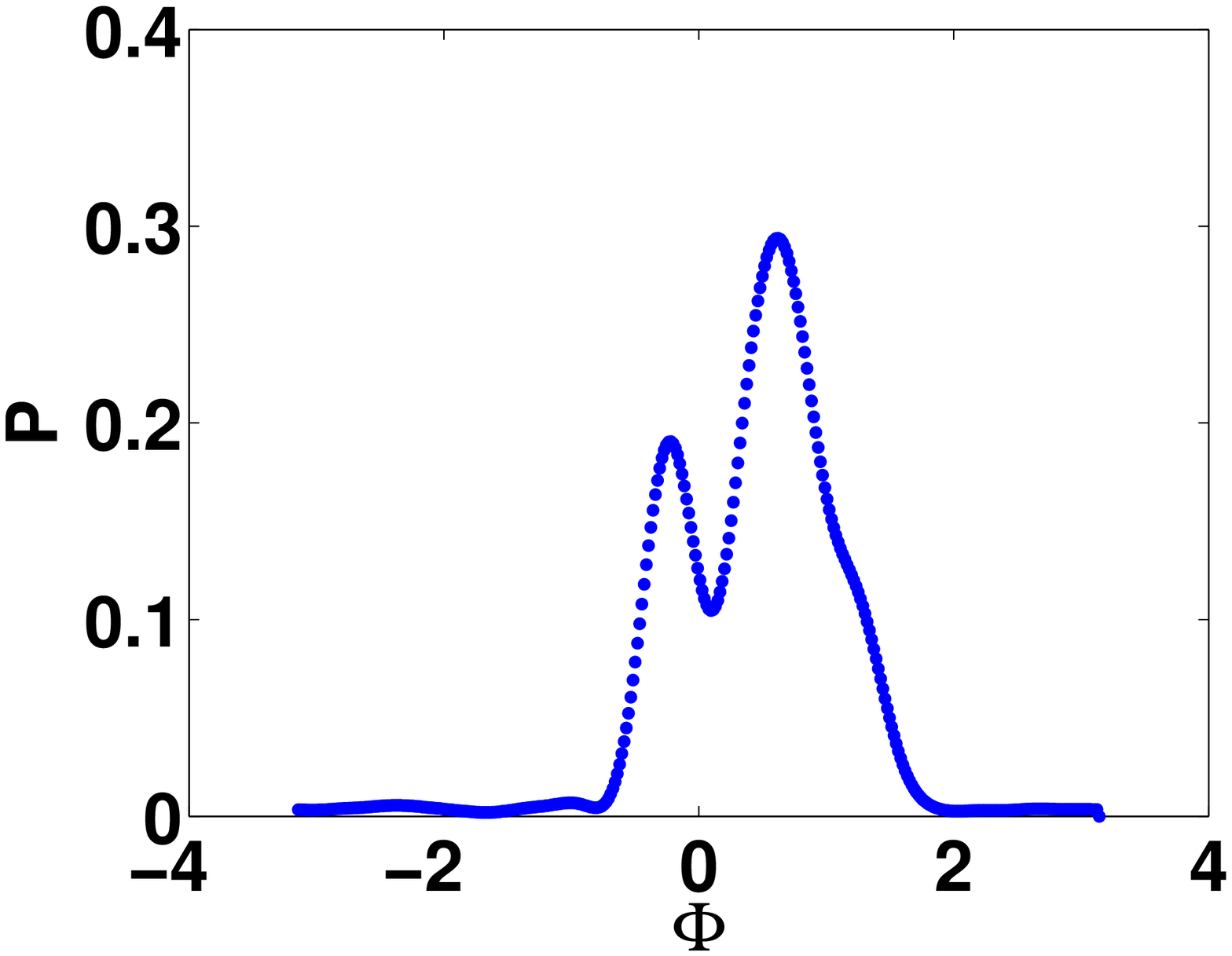}
\label{fig5b}}
\qquad
\subfloat[][]{
\includegraphics[width=0.24\textwidth]{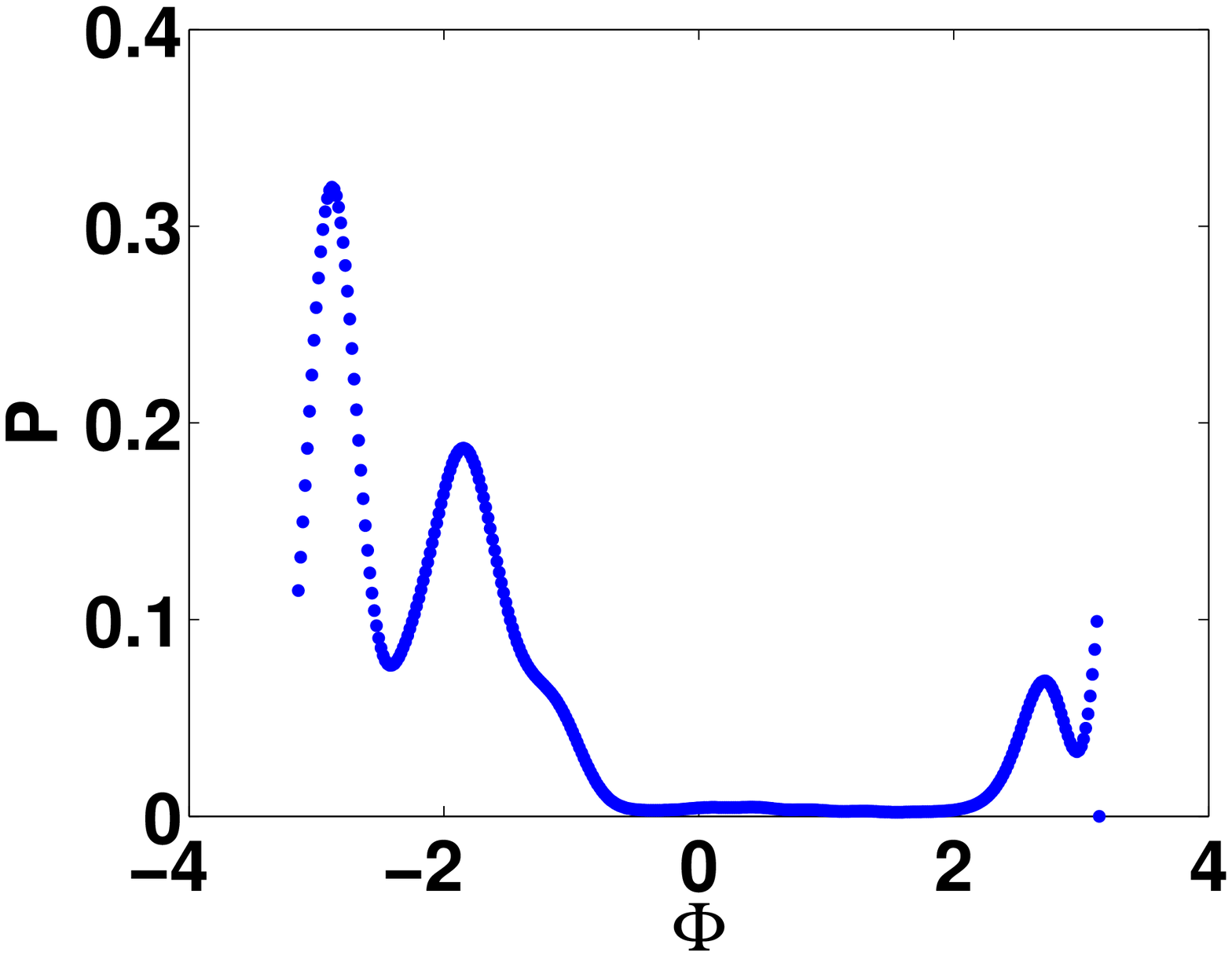}
\label{fig5c}}
\subfloat[][]{
\includegraphics[width=0.24\textwidth]{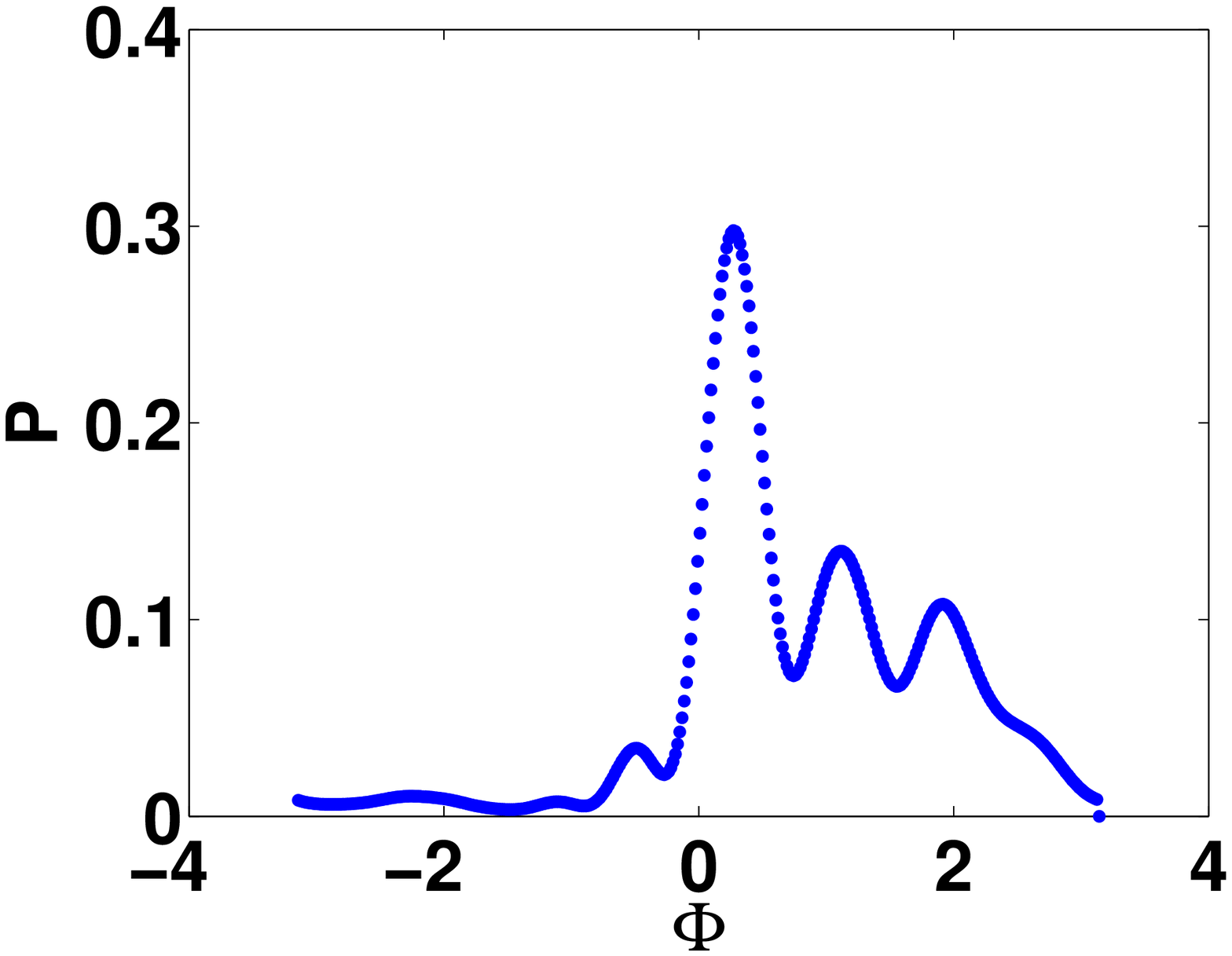}
\label{fig5d}}
\caption{
\label{fig5}
(Color online) The phase distribution $P=P(\Phi)$ as a function of the
angular variable $\Phi$ of the quasimagnon phase space  is shown
at the end of the (a) first, (b) second, (c) third, and the (d) fourth steps of the QW.
The initial phase distribution, wherein the QW starts at $\Phi=0$, is not shown.
The parameters are the same as Fig. \ref{fig2}. $\Phi$ is in radians.}
\end{figure}

\begin{figure*}[t]
\subfloat[][]{
\includegraphics[width=0.24\textwidth]{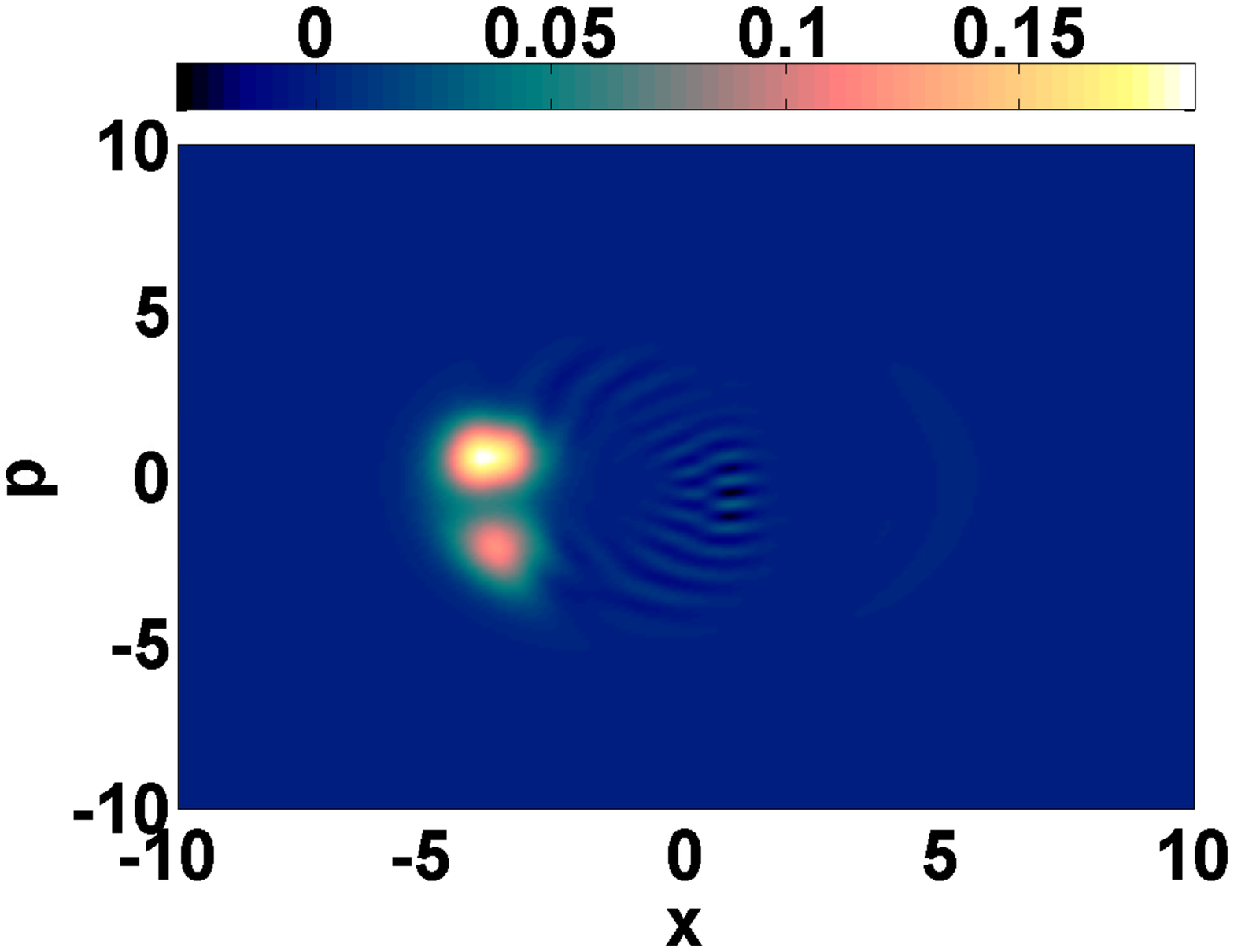}
\label{fig6a}}
\subfloat[][]{
\includegraphics[width=0.24\textwidth]{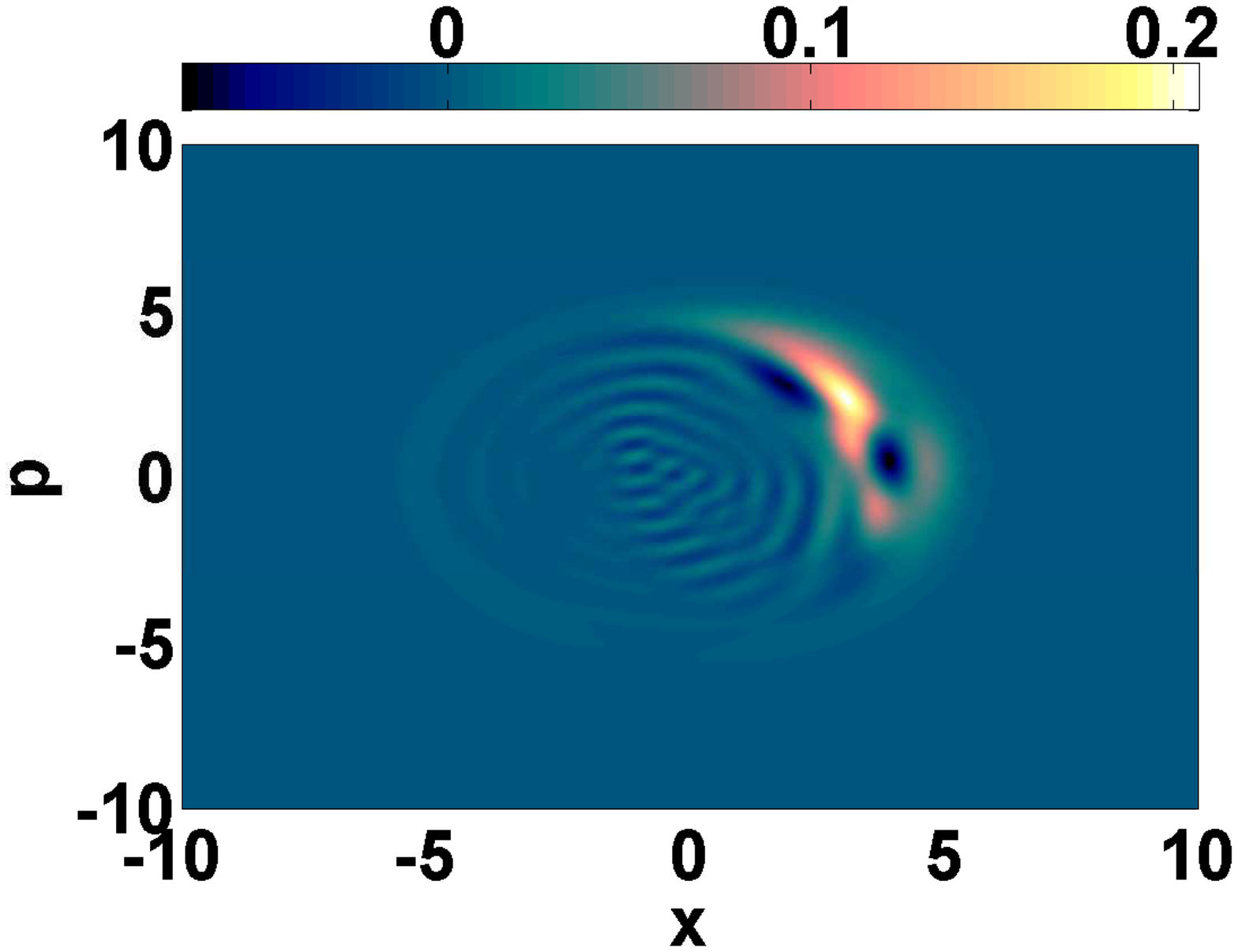}
\label{fig6b}}
\subfloat[][]{
\includegraphics[width=0.24\textwidth]{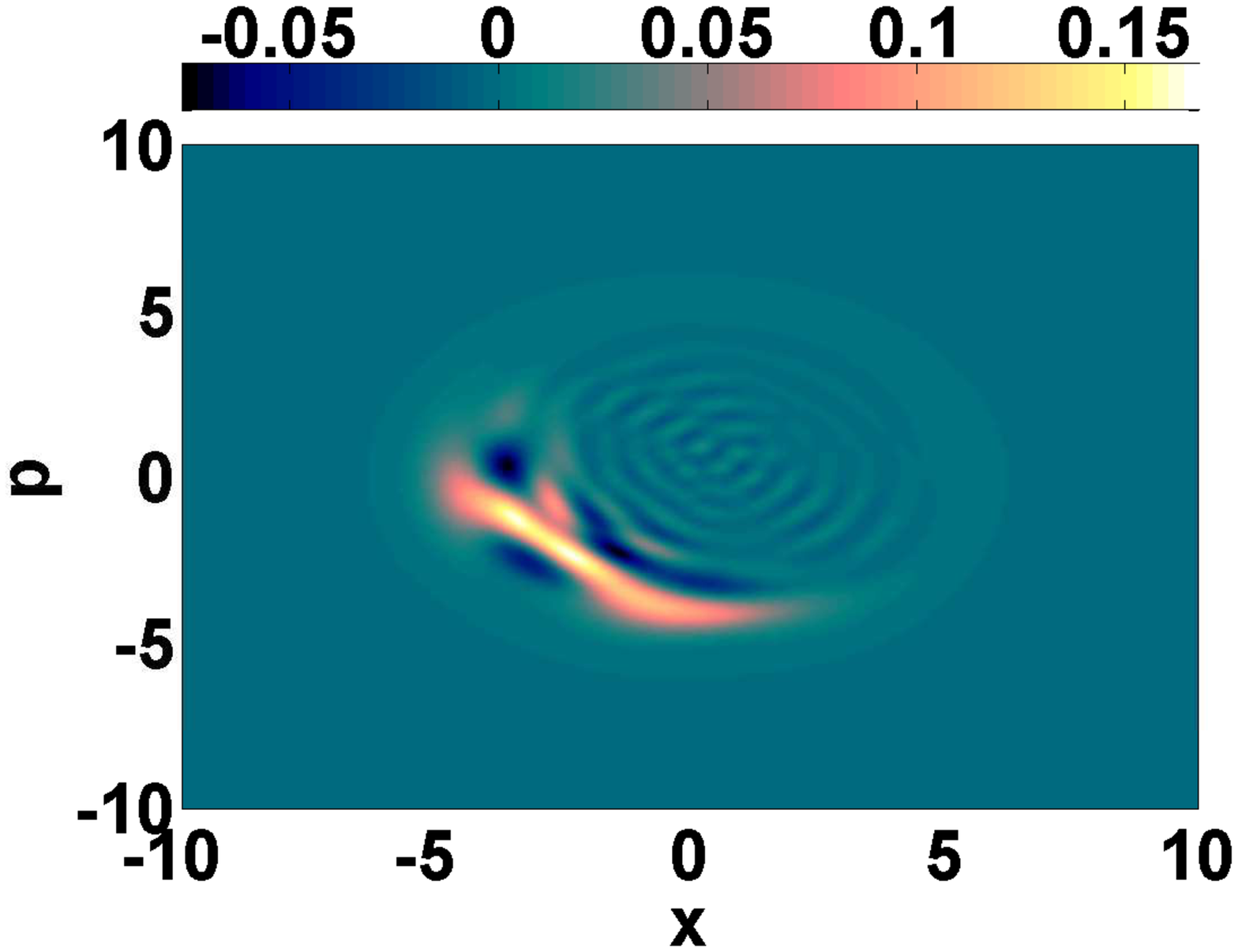}
\label{fig6c}}
\subfloat[][]{
\includegraphics[width=0.24\textwidth]{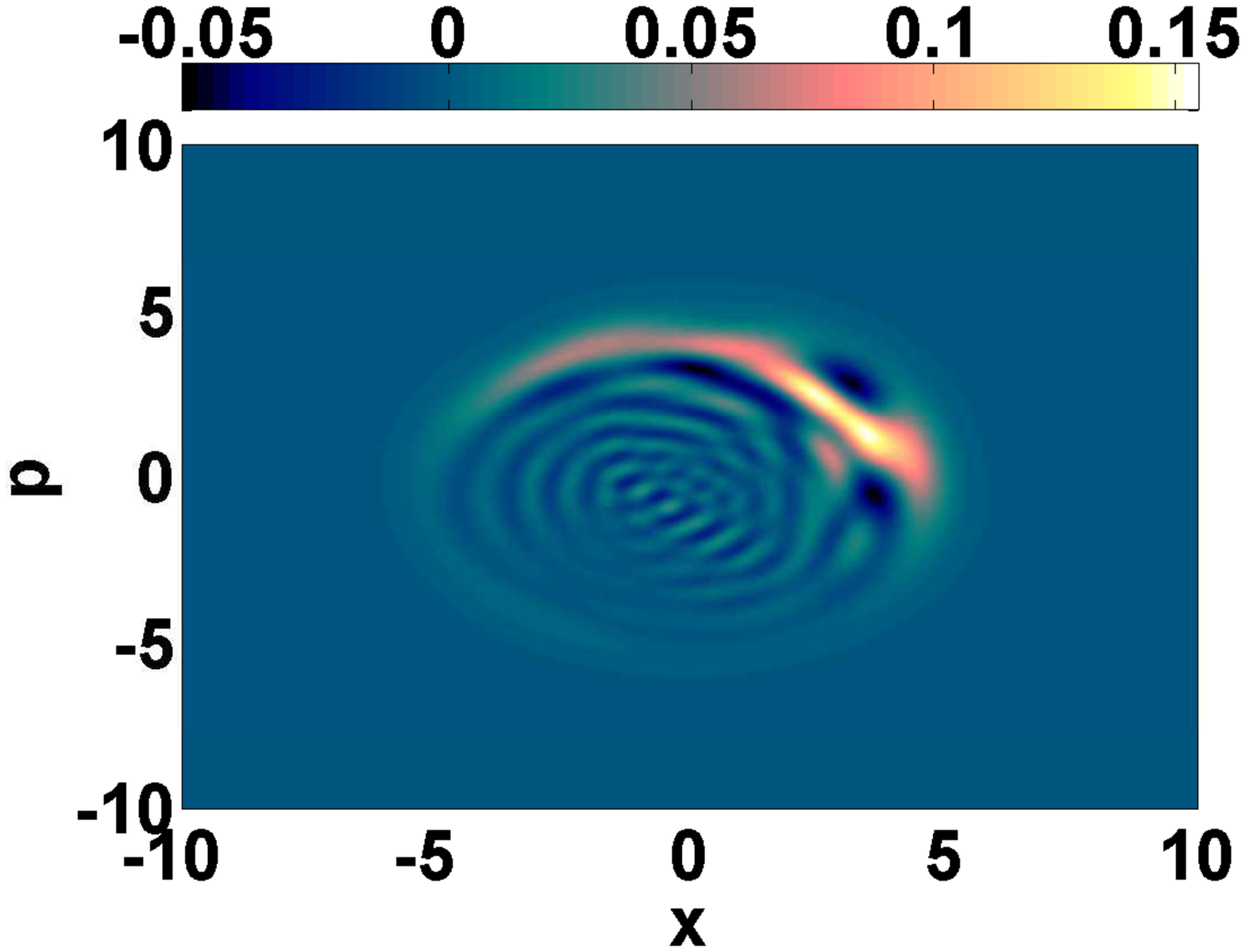}
\label{fig6d}}
\qquad
\subfloat[][]{
\includegraphics[width=0.24\textwidth]{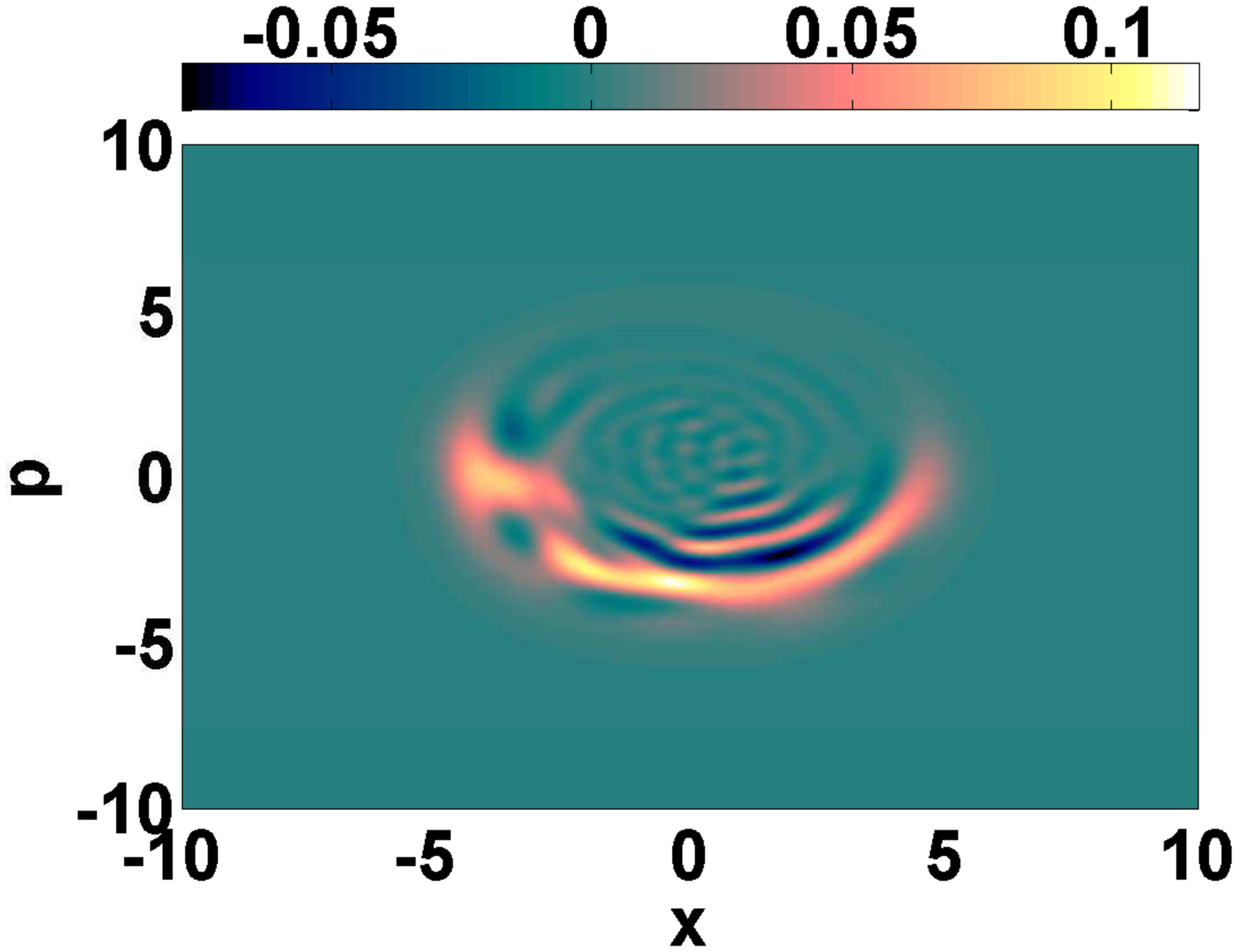}
\label{fig6e}}
\subfloat[][]{
\includegraphics[width=0.24\textwidth]{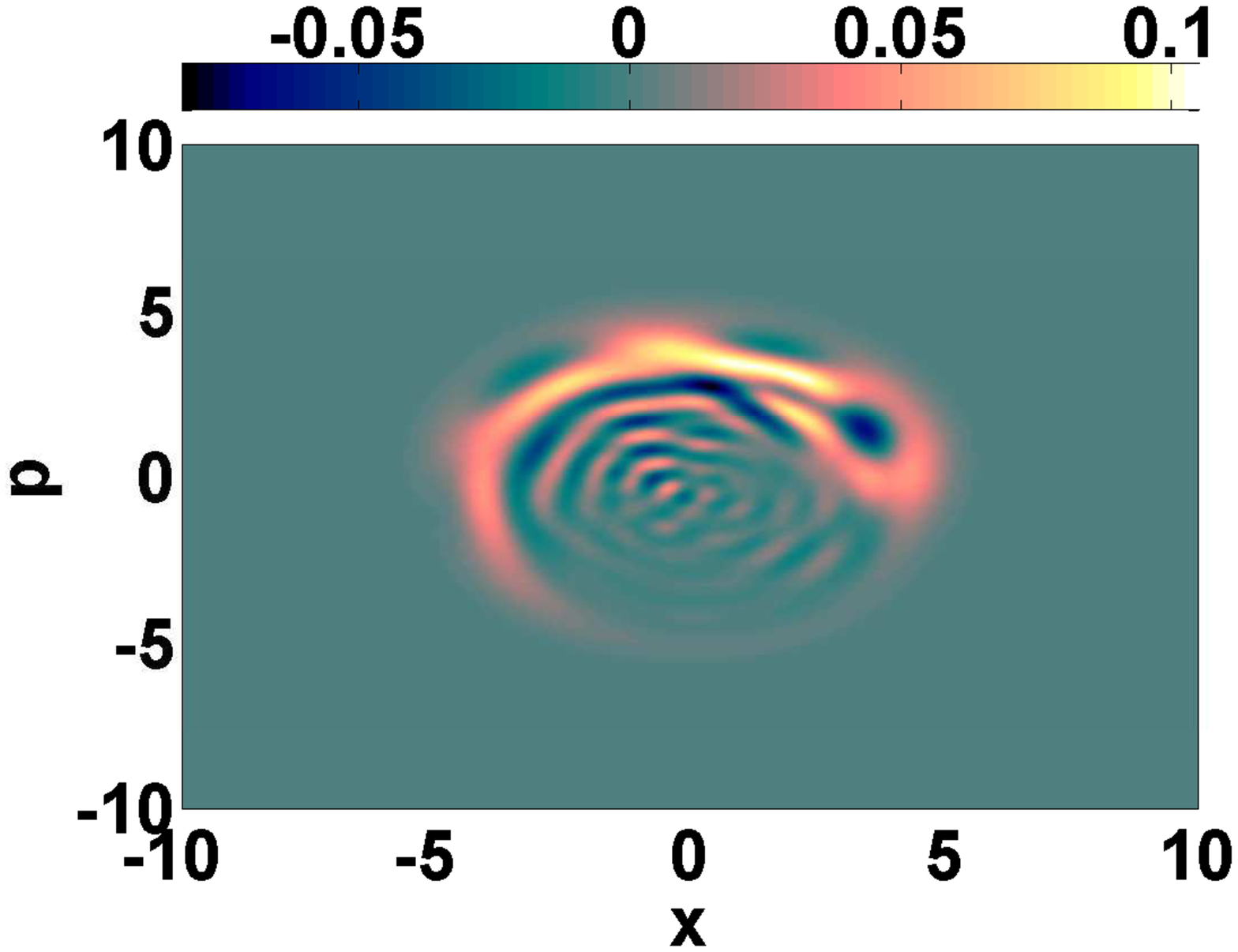}
\label{fig6f}}
\subfloat[][]{
\includegraphics[width=0.24\textwidth]{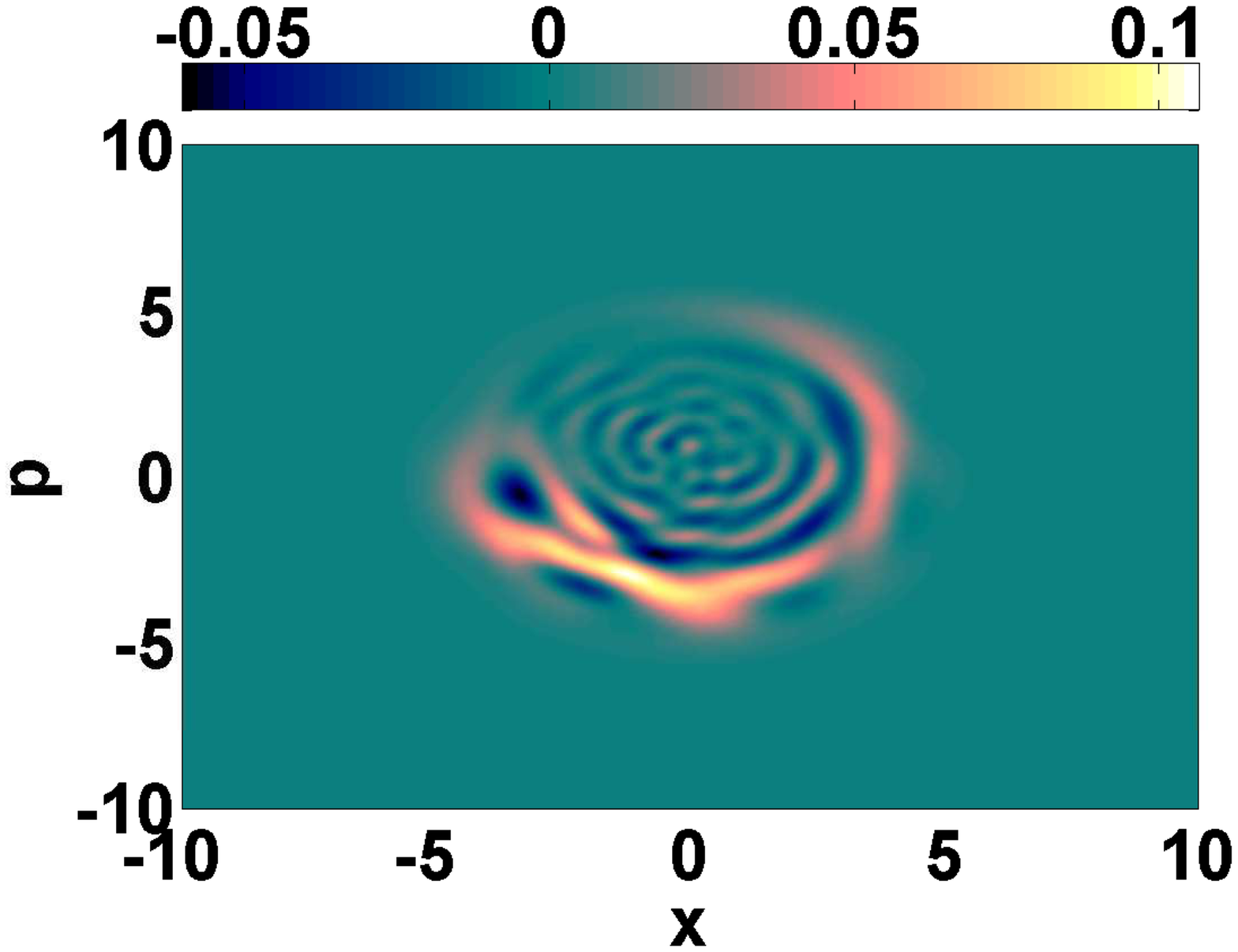}
\label{fig6g}}
\subfloat[][]{
\includegraphics[width=0.24\textwidth]{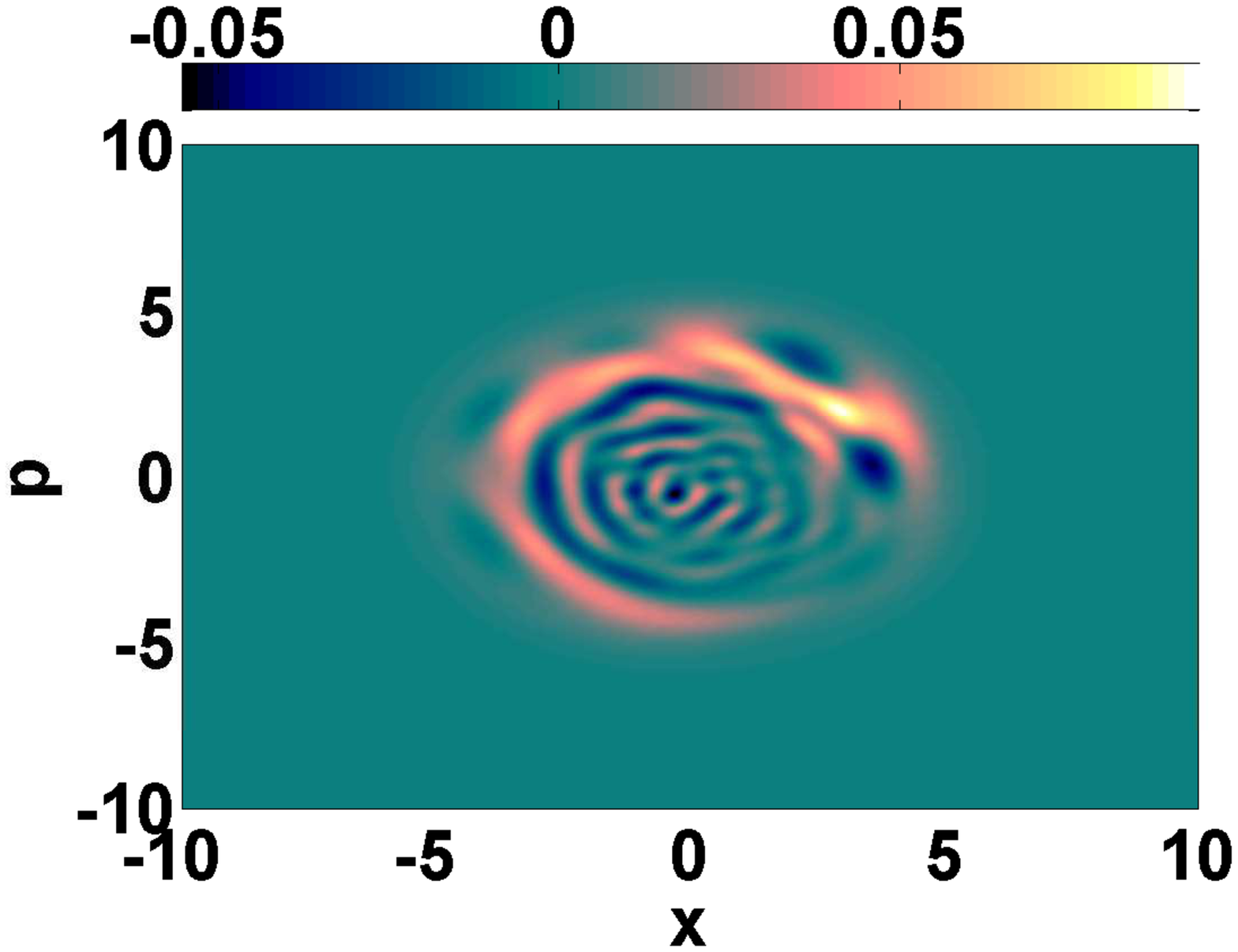}
\label{fig6h}}
\caption{
\label{fig6}
(Color online) The Wigner function at the end of the first eight steps of the QW.
The parameters are the same as Fig.~\ref{fig2}.
Real and imaginary parts of the coherent state parameter in the quasimagnon phase space
are denoted by $x$ and $p$, respectively.
The initial Wigner function, which is a Gaussian centered at $(x=3, p=0)$, is not shown.
The radius of the circle of the QW is determined by the square root of the mean number of the quasimagnons. Phase diffusion along
the angular direction is accompanied by the ripples along the radial direction due to the
number fluctuations as the QW progress.}
\end{figure*}

The observation of further QW signatures is to be expected, similar to earlier studies on the same Hamiltonian for similar parameters.
For example, the phase distributions at the end of the first four steps are reported in Ref.~\cite{xue2008}. Interpretation
of the phase distribution is not trivial because of the effects of the pump field. It is found that as the pump intensity grows, the
distribution becomes asymmetrically skewed in one direction. Moreover, some peaks become narrower.
We present the phase distribution in Fig.~\ref{fig5}. We used $M=256$ phase states in our numerical
calculations.
Phase distribution is skewed counterclockwise because of the pumping field, consistent with the
observations reported in Ref.~\cite{xue2008}. Because the walk occurs on a circle, the left and the right boundaries $\phi=\pm\pi$
in Fig. \ref{fig5} are the same. The initial phase distribution of the QW starts at an angular location of $\phi=0$. The spread of the distribution is quadratically enhanced, relative to that of a classical random walk.

Wigner functions at the end of the first eight steps are shown in Fig.~\ref{fig6}. The spread along the angular direction as the QW progress is accompanied by the increasing deformation along the radial direction due to the number fluctuations.
QW splits into other circles at different mean number of quasimagnons and dephasing occurs. Holevo measure increases beyond $1$ as the distribution gets more and more flat due to significant contribution of the phase diffusion after $7$ steps. Thus we did not use the Holevo measure to characterize the QW for more than $7$ steps.

We now examine the effect of decoherence rate of presently available NV center
ensembles with $\Gamma/2\pi=2.78$ MHz \cite{balasubramanian2009} which is about $10$ times larger than the
photons in cirQED and CQED QW proposals.
\begin{figure}[!t]
\begin{center}
\includegraphics[width=0.45\textwidth]{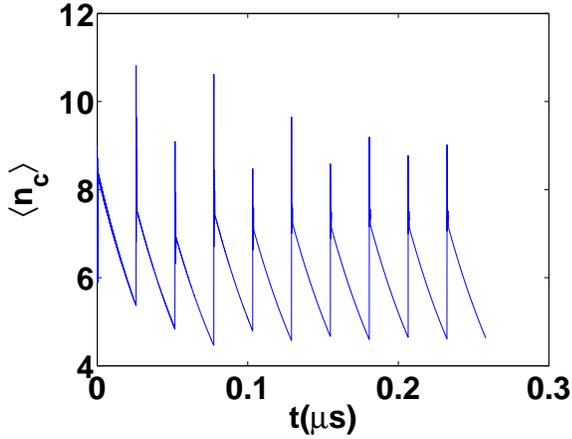}
\caption{\label{fig7} (Color online) Time (t) dependence of the mean quasimagnon number $\langle n_c\rangle$
under the same conditions as in Fig.~\ref{fig2} but for the
case of $\Gamma/2\pi=2.78$ MHz and drive
$\epsilon_0/2\pi=10$ GHz.}
\end{center}
\end{figure}
To compensate faster loss of quasimagnons, we take
$10$ times stronger drive with
$\epsilon_0/2\pi=10$ GHz in the simulations, which is found to be sufficient enough to stabilize the
circular trajectory of quantum walkers in the phase space by maintaining $\langle n_c\rangle \sim 6$
as can be seen in Fig. \ref{fig7}. For this drive strength, the drive duration becomes
$10$ times shorter, due to the relation $t_H\sim 1/\Omega_R\sim 1/\epsilon_0$.
The spikes in the figure correspond to the times when the drive is on or the action of coin toss. The Holevo standard
deviation shown in the Fig.~\ref{fig8}. In this case phase diffusion becomes significant after
four steps. When we determine the linear fit in logarithmic scale we find the
similar behavior as in Fig.~\ref{fig4} with the slope $0.98\pm 0.42$, where $0.42$ is the
linear-regression standard deviation for the first four steps.
Such a few number of steps yields large linear-regression standard deviation error in the slope.
The slope is nevertheless close enough to be declared to be ballistic,
given that phase diffusion due to significantly large
decoherence and wrapping due to periodicity of phase
have occured.

\begin{figure}[t]
\begin{center}
\includegraphics[width=0.45\textwidth]{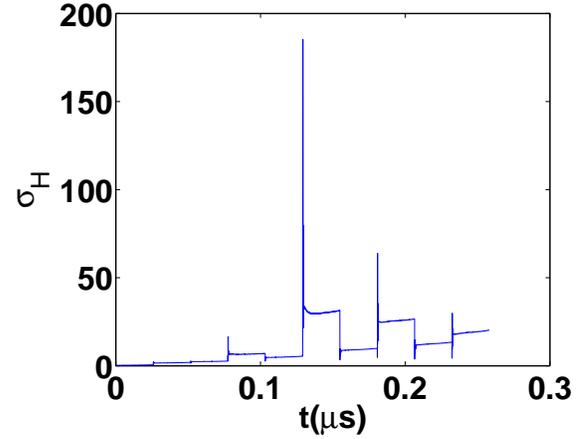}
\caption{\label{fig8} (Color online)  Time (t) dependence of the Holevo standard deviation measure $\sigma_\text{H}$
under the same conditions as in Fig.~\ref{fig3} but for the
case of $\Gamma/2\pi=2.78$ MHz and drive
$\epsilon_0/2\pi=10$ GHz.}
\end{center}
\end{figure}
\section{Conclusions}\label{sec:con}
We proposed a fully solid-state implementation of a DTQW of
NV centers in diamond coupled to a flux qubit. The QW takes place in the phase space of the quasimagnon field
by indirectly flipping the flux-qubit coin. The proposed scheme
is an improved version of the one originally developed for cirQED and cavity QED systems. In our method, the time steps are of fixed duration and no alteration of the pulse is required.
This improved method
is applicable to general cirQED and cavity QED QW scenarios. We did not find any significant reduction in the quality of the QW,
characterized by the spread of the Holevo standard deviation.

If the NV center ensemble is sufficiently large and is weakly excited, its excitations can be treated as bosonic quasiparticles, called quasimagnons. A driven Jaynes-Cummings model
for the quasimagnons and the flux qubit system is established. Further canonical transformation of the Jaynes-Cummings Hamiltonian to a QW on a cycle Hamiltonian is then applied, following a procedure similar to that
 used for a QW in cirQED. Calculation of the Holevo measure, by taking into account the dissipation channels, and then fitting a curve to it on a logarithmic scale, reveal that the spread of the
QW is quadratically faster than that of a classical random walk. Further, QW signatures are explored in the phase distribution and the Wigner function, and features that are similar to those found in the cirQED and cavity QED cases are confirmed.

\begin{acknowledgements}
The authors thank William J. Munro, Yuichiro Matsuzaki, Kousuke Kakuyanagi, and Shiro Saito for helpful discussions. Y. S. acknowledges financial support from Grant-in-Aid for Young Scientists (25800181). \"{O}.~E.~M. acknowledges financial support from T\"UB\.{I}TAK (Grant. No. 112T049).
B.~C.~S. acknowledges financial support from AITF, NSERC, and CIFAR. P.~X. acknowledges
financial support from NSFC 11004029 and 11174052.
\end{acknowledgements}
\appendix
\section{HUBBARD, SCHWINGER AND QUASIMAGNON OPERATORS}
\label{appendix:HubbardAlgebra}
Here we summarize the algebraic properties and relations among the Hubbard operators, bilinear
Schwinger boson forms and the quasimagnon operators. For notational simplicity we drop the NV center
crystallographic class index $f$. Let us introduce
the Gell-mann isotopic spin representation of SU(3)~\cite{gell-mann1953} in terms of the Hubbard operators,
\begin{eqnarray}
V_-&=&\frac{X^{0-}}{\sqrt{N}},\quad U_-=\frac{X^{0+}}{\sqrt{N}},\quad T_-=X^{-+},\\
V_3&=&\frac{X^{--}-X^{00}}{2N},\quad U_3=\frac{X^{++}-X^{00}}{2N},\\
T_3&=&\frac{X^{++}-X^{--}}{2},
\end{eqnarray}
with $V_+=V_-^\dag,U_+=U_-^\dag$ and $T_+=T_-^\dag$. Each $U,V,T$ spin-$1/2$ subgroup obeys SU(2) algebra,
$[J_-,J_+]=-2J_3,\,[J_3,J_{\pm}]=\pm J_{\pm}$ with $J=U,V,T$; but they are all correlated according to the
full set of commutation relations of the Lie algebra of SU(3).

SU(3) commutation relations can be obtained using the Hubbard
operator algebra
\begin{eqnarray}
[X^{mn},X^{m^\prime n^\prime}]=\delta_{m^\prime n}X^{m n^{\prime}}-\delta_{m n^\prime}X^{m^\prime n},
\end{eqnarray}
which is preserved under the bilinear Schwinger map
$X^{mn}=x_m^\dag x_n$. We associate the collective spin Hubbard operators
introduced for the NV center spin ensemble
with the Schwinger map subject to weak excitation condition
relative to large $N$ so that $x_0^\dag x_0\approx N\gg 1$.
In particular the $U$ and $V$
spins become decoupled
\begin{eqnarray}
[V_-,U_+]&=&-\frac{T_+}{N}\approx 0,
\end{eqnarray}
and contracted into Weyl-Heisenberg algebra according to
\begin{eqnarray}
[V_-,V_+]&=&-2V_3\approx 1.
\end{eqnarray}

The bosonization of the $U$ and $V$ spins in terms of the quasimagnon operators
\begin{eqnarray}
V_-&=&\frac{X^{0-}}{\sqrt{N}}=\frac{x_0^\dag x_-}{\sqrt{N}}\approx x_-\equiv b,\\
U_-&=&\frac{X^{0+}}{\sqrt{N}}=\frac{x_0^\dag x_+}{\sqrt{N}}\approx x_+\equiv a,
\end{eqnarray}
physically
means that the particle label counting the individual (NV center) spins in the collective spin has lost its meaning
for weak excitations in a large ensemble. This allows for quasimagnon description of weak excitations and spin
liquid interpretation of NV center ensemble. Mathematicaly this construction is equivalent to contraction of
$SU(3)$ into the semidirect product $SU(2)\bar\otimes h_2$ of $T$ spin algebra, which remains unaffected by the
bosonization condition, and quasimagnon Weyl-Heisenberg algebra~\cite{sun2003}.
\section{COUPLED AND UNCOUPLED QUASIMAGNON MODES WITH A FLUX QUBIT}
\label{appendix:quasimagnonModes}
Two quasimagnon modes corresponding to two quasispin waves with
different polarizations are nearly degenerate. This allows for decoupling
a combination of them from the flux qubit. The annihilation operators
for the coupled and uncoupled modes are $c_f=(a_f+b_f)/\sqrt{2}$
and $d_f=(a_f-b_f)/\sqrt{2}$, respectively. These are also nearly degenerate, whereby subsequent
decouplings are possible. New annihilation operators, given by
\begin{eqnarray}
c_{12}&=&\frac{1}{\sqrt{N_1+N_2}}(\sqrt{N_1}c_1+\sqrt{N_2}c_2),\nonumber\\
d_{12}&=&\frac{1}{\sqrt{N_1+N_2}}(\sqrt{N_2}c_1-\sqrt{N_1}c_2),\nonumber\\
c_{34}&=&\frac{1}{\sqrt{N_3+N_4}}(\sqrt{N_3}c_3+\sqrt{N_4}c_4),\nonumber\\
d_{34}&=&\frac{1}{\sqrt{N_3+N_4}}(\sqrt{N_4}c_3-\sqrt{N_3}c_4),\nonumber\\
\end{eqnarray}
are defined to decouple $d_{12}$ and $d_{34}$ from the qubit.

When the procedure is applied once more, it leads to the coupling of the qubit to a mode described by the annihilation operator
\begin{eqnarray}
c &=& \frac{\sqrt{N_1+N_2}c_{12}+\sqrt{N_3+N_4}c_{34}}{\sqrt{N}}\nonumber\\
&=&\frac{1}{\sqrt{2N}}(\sum_f\sqrt{N_f}(a_f+b_f),
\end{eqnarray}
with $N=\sum_fN_f$ as the total number of NV$^-$ centers. The uncoupled mode is given by
\begin{eqnarray}
d = \frac{\sqrt{N_3+N_4}c_{12}-\sqrt{N_1+N_2}c_{34}}{\sqrt{N}}.
\end{eqnarray}
\section{INHOMOGENEOUS COUPLING AND BROADENING OF NV CENTERS}
\label{appendix:inhomogenity}
Our model was developed under the assumptions of spatially homogeneous coupling of the NV center ensemble
to the flux qubit and absence of inhomogeneous broadening. Here we consider the more general case of
spatial inhomogeneity for which the model Hamiltonian $H=H_{\text{NV}}+H_q+H_{\text{int}}$ should be generalized
for the position dependent zero field splitting  $D_i$ and spatially inhomogeneous coupling $g_i$.

The NV center
ensemble Hamiltonian becomes
\begin{eqnarray}
H_{\text{NV}}=\sum_{f=1}^{4}\sum_{i=1}^{N_f}\sum_{m=0,\pm 1} D_im^2 X_{fi}^{mm},
\end{eqnarray}
where the Hubbard operators for each NV is defined to be
\begin{eqnarray}
 X_{fi}^{mm}=\mid m\rangle_i^{(f)}\langle m\mid_i^{(f)}.
 \end{eqnarray}

 For the single NV center the weak excitation condition can be used for introducing bilinear Schwinger map
 $X_{fi}^{mn}=x_{fim}^\dag x_{fin}$ with $x_{fi0}\approx 1$. Using the notation $x_{fi+}\equiv a_{fi}$ and
$x_{fi-}\equiv b_{fi}$, we write the $H_{\text{int}}$ as
\begin{eqnarray}
H_{\text{int}}=\sum_{fi}g_i[\sigma_+(a_{fi}+b_{fi})+\text{H.c.}].
\end{eqnarray}
Introducing $c_{fi}=(a_{fi}+b_{fi})/\sqrt{2}$ and dropping the uncoupled mode $d_{fi}=(a_{fi}-b_{fi})/\sqrt{2}$,
the Hamiltonian becomes a spin-boson model
\begin{eqnarray}
H=\frac{\omega_q}{2}+\sum_{fi}D_ic_{fi}^\dag c_{fi}+g_i(\sigma_+c_{fi}+\text{H.c.}).
\end{eqnarray}

We can now identify the bosonic collective mode that is coupled to the flux qubit as
\begin{eqnarray}
c=\frac{1}{G}\sum_fG_fc_f,
\end{eqnarray}
where
\begin{eqnarray}
G_f=\sqrt{2\sum_i g_i^2},\quad G=\sqrt{\sum_f G_f^2},
\end{eqnarray}
and
\begin{eqnarray}
c_f=\frac{1}{G_f}\sum_i g_i (a_{fi}+b_{fi}).
\end{eqnarray}
This shows that quasimagnon picture can be used in the presence of spatially dependent coupling;
albeit the quasimagnon mode would be subject to dephasing effect of the inhomogeneous broadening.

The coupling $H_{\text{int}}=G(\sigma_+ c+\text{H.c.})$ of the flux qubit and the quasimagnon mode
leads to a qubit-polariton energy gap. This gap can protect the quasimagnon against dephasing if it is
sufficiently strong. A more intriguing proposal would be to exploit a particular inhomogeneous coupling density profile
$\rho(\omega)$
to optimize linewidth reduction~\cite{kurucz2011}.
This method has been suggested for cavity-spin ensemble coupling~\cite{kurucz2011}, where it has been shown that
a Gaussian coupling density profile reduce the polariton linewidth to a limit independent of inhomogeneous linewidth.
In general, it has been found that  the coupling density should fall off as $1/\omega^3$ or
faster for the manifestation of the polariton gap induced line narrowing.

 Another challenge for the experimental implementation of our model in the presence of spatial inhomogeneity would be
 to drive the exact quasimagnon mode. For spatially homogeneous case this mode is always accessed by the drive. If the drive field
 spatial profile can be adjusted to match with the magnetic field profile of the flux qubit that couples to the NV center
 ensemble, then the quasimagnon mode in the spatial inhomogenity case can be driven as well.

\end{document}